\documentclass[12pt,draftclsnofoot,onecolumn]{IEEEtran}
\usepackage{xcolor}
\usepackage{balance}

\usepackage{cite}
\usepackage{amsmath,amssymb,amsfonts}
\usepackage{graphicx}
\usepackage{textcomp}
\usepackage{acronym}
\usepackage{xcolor}
\usepackage{tikz} % for tikz
\usepackage[utf8]{inputenc}
\usepackage{pgfplots} 
\usepackage{pgfgantt}
\usepackage{pdflscape}
\usepackage{changes}
\usepackage{comment}
\usepackage{subfigure}
\usepackage{mathtools,algpseudocode,algorithm,MnSymbol}
\usepackage{algorithm}
\usepackage{pgfplots}
  \pgfplotsset{compat=newest}
  \usetikzlibrary{plotmarks}
  \usetikzlibrary{arrows.meta}
  \usepgfplotslibrary{patchplots}
  \usepackage{grffile}
  \usepackage{amsmath}

\pgfplotsset{compat=newest} 
\pgfplotsset{plot coordinates/math parser=false} % end of tikz

\acrodef{rbp}[RBP]{Rao-Blackwellized particle}
\acrodef{rbpf}[RBPF]{RBP filter}
\acrodef{ue}[UE]{user equipment}
\acrodef{bs}[BS]{base station}
\acrodef{tomb}[TOMB/P]{track-oriented  marginal multi-Bernoulli/Poisson}
\acrodef{fov}[FoV]{field-of-view}   
\acrodef{momb}[MOMB/P]{measurement-oriented  marginal multi-Bernoulli/Poisson}

\newtheorem{example}{Example}
\begin{document}

\bibliographystyle{IEEEtran}
\bstctlcite{IEEEexample:BSTcontrol}
%\title{An End-to-end 5G SLAM Framework with a Low-complexity Channel Estimator}
\title{A Computationally Efficient EK-PMBM Filter for Bistatic mmWave Radio SLAM}% in  Sensing Applications}

\author{
Yu Ge,~\IEEEmembership{Student~Member,~IEEE,}   % 1st author, 1st affiliations
Ossi Kaltiokallio,
Hyowon Kim,~\IEEEmembership{Member,~IEEE,} \\
Fan Jiang,~\IEEEmembership{Member,~IEEE,} 
Jukka Talvitie,~\IEEEmembership{Member,~IEEE,} 
Mikko Valkama,~\IEEEmembership{Senior~Member,~IEEE,}
Lennart Svensson,~\IEEEmembership{Senior~Member,~IEEE,} 
Sunwoo Kim,~\IEEEmembership{Senior~Member,~IEEE,}
Henk Wymeersch,~\IEEEmembership{Senior~Member,~IEEE}     % 4th author, 4th affiliations
\thanks{Yu Ge, Fan Jiang, Lennart Svensson and Henk Wymeersch are with the Department of Electrical Engineering, Chalmers University of Technology, Gothenburg, Sweden. Emails: 
\{yuge,fan.jiang,lennart.svensson,henkw\}@chalmers.se.}
\thanks{Ossi Kaltiokallio, Jukka Talvitie and Mikko Valkama are with the Unit of Electrical Engineering, Tampere University, Tampere, Finland. Emails: \{ossi.kaltiokallio,jukka.talvitie,mikko.valkama\}@tuni.fi}
\thanks{Hyowon Kim and Sunwoo Kim are with the Department of Electronic Engineering, Hanyang University, Seoul, South Korea. Email: \{khw870511,remero\}@hanyang.ac.kr.}
\thanks{This work was partially supported by the Wallenberg AI, Autonomous Systems and Software Program (WASP) funded by Knut and Alice Wallenberg Foundation, and the Vinnova 5GPOS project under grant 2019-03085, by the Swedish Research Council under grant 2018-03705.}
}



\maketitle

\begin{abstract}
Millimeter wave (mmWave) signals are useful for simultaneous localization and mapping (SLAM), due to their inherent geometric connection to the propagation environment and the propagation channel. To solve the SLAM problem, existing approaches rely on sigma-point  or  particle-based approximations, leading to high computational complexity, precluding real-time execution. We propose a novel low-complexity SLAM filter, based on the Poisson multi-Bernoulli mixture (PMBM) filter. It utilizes the extended Kalman (EK) first-order Taylor series based Gaussian approximation of the filtering distribution, and applies the \ac{tomb} algorithm to approximate the resulting PMBM as a Poisson multi-Bernoulli (PMB). The filter can account for different landmark types in radio SLAM and multiple data association hypotheses. Hence, it has an adjustable complexity/performance trade-off. 
Simulation results show that the developed SLAM filter can greatly reduce the computational cost, while it keeps the good performance of mapping and user state estimation.
\end{abstract}

\vskip0.5\baselineskip

% \fi
% \IEEEpeerreviewmaketitle

\section{Introduction}
5G and Beyond 5G systems can provide high-resolution measurements of delays and angles, which make them attractive for localization and sensing applications \cite{wymeersch20175g, witrisal2016high, kanhere2021target, zhang2021overview}. Localization of connected devices is important for autonomous vehicles \cite{reid2019localization},  Vehicle-to-Everything (V2X) \cite{ko2021v2x,zhang2020cooperative}, and spatial signal design \cite{furkan2021optimal}. Sensing of passive objects is important for joint radar and communication \cite{luong2021radio,kumari2021jcr70,han201224}. 
Sensing can be monostatic or bistatic, where in monostatic sensing (e.g., in automotive radar) the transmitter and receiver are co-located, which brings the advantage of a common clock and perfect knowledge of the transmitted data signal \cite{reichardt2012demonstrating}. %Even when the radar's global location is unknown, the backscattered signal provides valuable information about the environment, in the frame of reference of the radar. On the other hand, the monostatic operation required full-duplex transceivers, which must operate with over 100 dB signal power difference between the strong transmitted signal and the weak backscattered signal. 
In bistatic sensing, the transmitter and receiver are spatially separated \cite{kanhere2021target}. Hence, the data symbols are unknown to the receiver, and the transmitter and receiver are not synchronized. The former issue can be resolved by sending predetermined pilot signals (in time, frequency, and space), while the latter issue has the serious implication that only delay-differences among propagation paths bring information. 
When the transmitter and receiver have known locations (e.g., \acp{bs}), the problem is referred to as passive localization or mapping. When the transmitter or the receiver has an unknown location (e.g., a \ac{ue}), this location must be determined jointly with the map (this also applies to the monostatic case with a \ac{ue} radar), referred to as radio simultaneous localization and mapping (SLAM) \cite{durrant2006simultaneous,bailey2006simultaneous}. Here, the \ac{ue} acts as a \emph{sensor} with an unknown and time-varying state, while the static objects in the propagation environment act as \emph{landmarks} with unknown states and cardinality. 

Solving the radio SLAM problem is challenging because i) channel estimation errors, or noise peaks may result in false detections, ii) landmarks in the \ac{fov} of the sensor
can be undetected, due to the imperfect detection performance at the sensor, iii) the number of landmarks in the \ac{fov} is primarily unknown, as the map is unknown, iv) the source (landmark) of each measurement is unknown at the sensor, so there is an inherent data association problem \cite{bar1990tracking}. 
%SLAM is a classical topic in robotics and there have been 
Several approaches have been proposed to address the above-mentioned challenges, including methods based on geometry \cite{yassin2018mosaic,aladsani2019leveraging}, on belief-propagation (BP)
\cite{Rico_BPSLAM_JSTSP2019,Erik_BPSLAM_TWC2019,Erik_AOABPSLAM_ICC2019},
% \cite{loeliger2004introduction, kschischang2001factor},
and on random-finite-set (RFS) theory \cite{mullane2011random, mahler2014advances}. Geometry-based methods \cite{yassin2018mosaic,aladsani2019leveraging} have low complexity, but cannot inherently deal with the unknown number of targets or the data association problem. 
% In \cite{mendrzik2018joint,mendrzik2018harnessing,kim20185g}, the BP-based estimators are developed, however, the data association problem is still not considered.
{BP-SLAM methods~\cite{Rico_BPSLAM_JSTSP2019,Erik_BPSLAM_TWC2019,Erik_AOABPSLAM_ICC2019} have been developed from BP multi-target tracking (MTT)~\cite{meyer2018message}, which extends the BP-based computation of marginal association probabilities from RFS-based approaches~\cite{williams2015marginal}. 
In BP-SLAM, the landmark state is modeled as a vector instead of an RFS, which requires ad-hoc modifications to deal with appearing and disappearing landmarks. Moreover, BP-SLAM cannot account for correlation between the sensor state and the landmark states directly, due to the marginalization carried out at each step.}
%Thus, the Poisson model for undetected landmarks is required to be considered as auxiliary PHD outside of the BP-SLAM framework.}
%Compared with geometry-based and BP-based methods,

RFS-based methods are particularly attractive for the SLAM problem, because the set of landmarks are modeled as an RFS,
% a random finite set,
where uncertainties on both cardinality and state of each landmark are considered, and there is no ordering of landmarks~{\cite{mahler2003multitarget}}. %. This makes the between the physical reality and the set of states as well as the data association problem more clear
Among RFS-based methods, the Poisson multi-Bernoulli mixture (PMBM) density is known to be a conjugate prior for most common measurement and dynamic models~{\cite{garcia2021poisson}}. %, and can thus be seen as a RFS equivalent of the Kalman filter for the linear and Gaussian vector case.
The key to the optimality of the PMBM filter lies in keeping track of all possible data association hypotheses over time, conditioned on the unknown sensor state, which renders the PMBM computationally demanding. To reduce the complexity, its simplified version, the Poisson multi-Bernoulli (PMB) filter, can be used, which reduces the number of hypotheses to one after each update. Several methods exist for this reduction, including the track-oriented  marginal multi-Bernoulli/Poisson (\ac{tomb}), the \ac{momb}~\cite{williams2015marginal}, and the Kullback-Leibler
divergence minimization \cite{williams2014efficient} algorithms. The PMBM family of filters has been used for MTT with known sensor state. To account for unknown sensor states, a low-complexity PMBM filter for joint MTT and sensor tracking is proposed in~\cite{frohle2019multisensor}
% , proposed a low-complexity PMBM filter for joint MTT and sensor tracking, 
by approximating the joint density with the product of marginal densities and performing separate updates per target. 

Several RFS-based methods have been proposed for mmWave radio SLAM \cite{kim20205g, kim2020joint,ge2020exploiting, ge20205GSLAM,EKPHD2021Ossi,HyowonPMBM}, which must account for the specific properties of mmWave sensing, including the highly nonlinear measurement models, time-varying detection probabilities, and the presence of multiple measurement models, due to different landmark types. These approaches mainly differ in terms of the representation of the RFS density and the required approximations. Among those RFS-based SLAM methods, \cite{kim20205g} developed a \ac{rbp} probability hypothesis density (PHD) filter. However, the complexity increases exponentially with the state  dimension. To reduce the complexity, \cite{kim2020joint} introduced a  cubature  Kalman  PHD  (CK-PHD) filter for radio-SLAM. Although the CK-PHD exhibits a lower computational cost than the RBP-PHD,
% Although CK-PHD is relatively low-cost, compared to RBPF-PHD, 
it relies on the sigma-point approximation,
% which cannot be executed in real-time. In order to further reduce the complexity,  extended  Kalman  PHD  (EK-PHD) was introduced in \cite{EKPHD2021Ossi}. 
and thus the computational cost can be further reduced by the extended Kalman PHD (EK-PHD) in \cite{EKPHD2021Ossi}.
These three methods are based on the PHD filter, which considers the data association problem, but there is no  explicit enumeration of different data associations. 
To explicitly consider all possible data associations, the RBP-PMBM SLAM filter is considered in \cite{ge2020exploiting, ge20205GSLAM}. The PMBM filter has better mapping performance than the PHD filter~\cite{HyowonPMBM}, through exhibiting high computational cost.
% The RBPF-PMBM SLAM filter considered in \cite{ge2020exploiting, ge20205GSLAM}. Because the PMBM filter enumerates all possible data associations explicitly, it outperforms the PHD filter, as shown in \cite{HyowonPMBM}, but has higher complexity. 

%In \cite{frohle2019multisensor}, a low-complexity PMBM filter is proposed for target tracking, but does not consider multiple measurements models present in mmWave SLAM.     used to track objects and the sensor, but the proposed algorithm cannot deal with the uncertainty of the measurement model, and the update is done landmark by landmark. therefore, the correction among landmarks is not considered.

%

In this paper, we address the high complexity of the RBP-PMBM filter by proposing a novel approach based on the extended Kalman (EK) filter, which performs a joint update of the sensor and
% map 
landmark states. Compared to the RBP-SLAM filters from \cite{kim20205g,ge2020exploiting,ge20205GSLAM}, the resulting filter has very low complexity and can thus be applied to real-time
%provide an efficient and effective PMB(M)-based SLAM filter for 
\ac{ue} localization and environment mapping in bistatic mmWave sensing with a single \ac{bs}. Compared to the EK-PHD \cite{EKPHD2021Ossi} filter, which accounts only for the most likely data association and landmark type, we show how to explicitly account for several data associations and multi-model (MM) implementation. 
In contrast to \cite{frohle2019multisensor}, the proposed filter can cope with multiple nonlinear measurement models, and applies a joint update to all landmarks and the sensor state. %Similar to \cite{EKPHD2021Ossi}, we list the top-ranked data associations but develop an update and PMB approximation, tailored to the PMBM density. 
%herefore, we proposed an extended Kalman filter (EKF)-PMB SLAM filter. The proposed method is built on the PMBM filter, utilizes the extended Kalman filter to jointly update the \ac{ue} and landmark states for each data association, and uses the TOMB algorithm \cite[Fig.\,10]{williams2015marginal} to marginalize over all data associations to keep the PMB format. 
Our main contributions are summarized as follows:
\begin{itemize}
    \item The derivation of the EK-PMBM SLAM filter, which uses a new and theoretically sound method to jointly update the \ac{ue} state and landmark states.  
    \item The derivation of the EK-PMB SLAM filter with a novel algorithm to approximate the resulting PMBM to a PMB, which is based on the TOMB algorithm with a limited number of data associations;
    \item The extension of the EK-PMBM and EK-PMB SLAM filters to the case of multiple landmark types, resulting in a multi-model (MM) implementation with hybrid discrete and continuous landmark states; % of the EK-PMB SLAM filter, which considers different measurement models, and can not only maps the environment but also distinguishes the landmark types;
    \item The validation of the proposed SLAM filters,   showing it exhibits very low complexity compared to the RBP-PMBM filter, while maintaining comparable SLAM performance. %as the RBP-PMBM SLAM filter.
\end{itemize}

The remainder of this paper is structured as follows. The system models are described in Section \ref{model}. The PMBM density and the Bayesian recursion of RFS-joint SLAM density are then introduced in Section \ref{PMBM_SLAM_Filter}. The novel EK-PMBM SLAM filter is derived in Section \ref{EKF-PMBM}, and its PMB counterpart in Section \ref{EKF-PMB}. The extension to multiple measurement models is covered in Section \ref{ekf-mm-pmbm}. Simulation results are presented in Section \ref{Results}, followed by our conclusions in Section \ref{conclusion}.

\subsubsection*{Notations} Scalars (e.g., $x$) are denoted in italic, vectors (e.g., $\boldsymbol{x}$) in bold, matrices (e.g., $\boldsymbol{X}$) in bold capital letters, sets  (e.g., $\mathcal{X}$) in calligraphic, and its cardinality is denoted as $\left|\mathcal{X}\right|$. Transpose is denoted by $(\cdot)^{\mathsf{T}}$. A Gaussian density with mean $\boldsymbol{u}$ and covariance $\boldsymbol{C}$, evaluated in value $\boldsymbol{x}$ is denoted by $\mathcal{N}(\boldsymbol{x};\boldsymbol{u},\boldsymbol{C})$. The union of mutually disjoint sets is denoted by $\uplus$, and  the Kronecker product is denoted by $\otimes$.

\section{System Models in mmWave Bistatic Sensing} \label{model}
\begin{figure}
    \centering
    \includegraphics[width=0.5\linewidth]{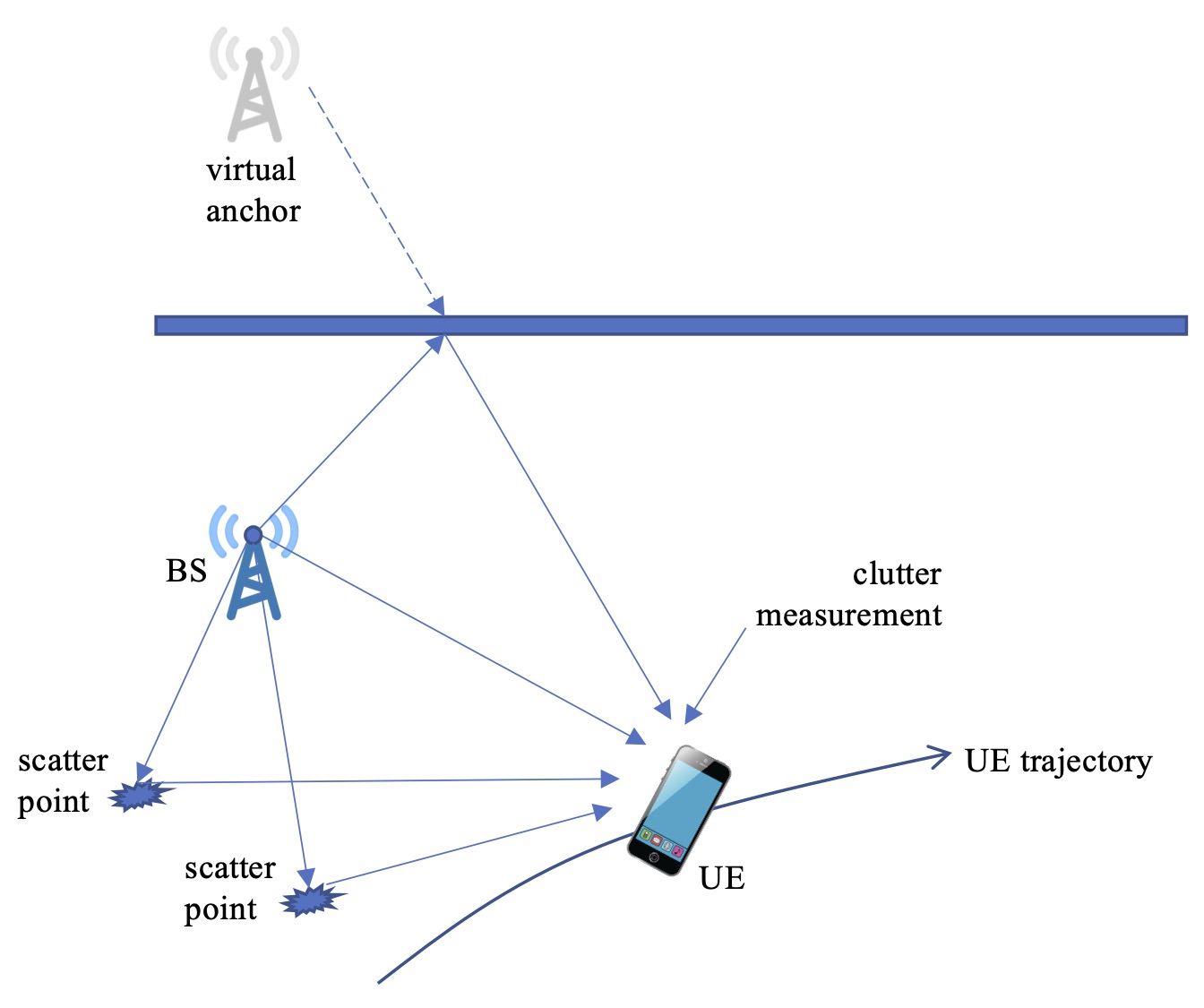}
    \caption{Bistatic mmWave sensing scenario, where the UE tracks its own state and constructs a map of the environment, by incorporating bistatic measurements from received downlink signals sent by the BS.}
    \label{fig:overview}
\vspace{-4mm} \end{figure}

    In this section, we introduce the models for the mobile \ac{ue} state and the state of the landmarks in the propagation environment of mmWave, as shown in Fig.~\ref{fig:overview}.
    Then, we provide the measurement model for the mmWave bistatic sensing scenario.
    
% While the proposed methods are applicable to general SLAM problems, our application is on mmWave bistatic sensing, as shown in Fig.~\ref{fig:overview}. 

\subsection{State Models} \label{State model}
We consider a multi-antenna  \ac{bs} with  known  location $\boldsymbol{x}_{\mathrm{BS}}\in \mathbb{R}^{3}$ and a  multi-antenna \ac{ue}, with sensor state at time step $k$ denoted by $\boldsymbol{s}_{k}$ (containing at least the 3D position and clock bias). The user dynamics are given by
%We model the process noise as a  zero-mean Gaussian, then the transition density of $\boldsymbol{s}_{k}$ given $\boldsymbol{s}_{k-1}$ can be expressed as
\begin{equation}
f(\boldsymbol{s}_{k+1} | \boldsymbol{s}_{k}) = {\cal N}(\boldsymbol{s}_{k+1} ; \boldsymbol{v}(\boldsymbol{s}_{k}),\boldsymbol{Q}_{k+1}), \label{dynamicmodel}
\end{equation}
where $\boldsymbol{v}(\cdot)$ denotes a known transition function and $\boldsymbol{Q}_{k+1}$ is the process noise covariance. %The above dynamic model can be expressed, equivalently, in terms of a transition density $f(\boldsymbol{s}_k | \boldsymbol{s}_{k-1}) = {\cal N}(\boldsymbol{s}_k ; v(\boldsymbol{s}_{k-1}),\boldsymbol{Q})$. 
The environment comprises three different types of landmarks, the \ac{bs}, scattering points (SPs) and  reflecting surfaces. Each SP is parameterized by an unknown 3D location $\boldsymbol{x}_{\mathrm{SP}}\in \mathbb{R}^{3}$, while each reflecting surface is parameterized by a fixed virtual anchor (VA) with location $\boldsymbol{x}_{\mathrm{VA}}\in \mathbb{R}^{3}$. The VA  is the reflection of the \ac{bs} with respect to the reflecting surface {\cite{Rappaport_6G100GHz_Access2019,palacios2019single}}
% kim20205g}:
\begin{align}
    \boldsymbol{x}_{\mathrm{VA}}=(\boldsymbol{I}-2\boldsymbol{\nu}\boldsymbol{\nu}^{\mathsf{T}}) \boldsymbol{x}_{\mathrm{BS}}+2\boldsymbol{\mu}^{\mathsf{T}}\boldsymbol{\nu} \boldsymbol{\nu},
\end{align}
where $\boldsymbol{\mu}$ is an arbitrary point on the surface, and $\boldsymbol{\nu}$ is the normal to the reflecting surface. The VA is surface-specific; although the incidence point of the downlink signal on the reflecting surface is moving while the \ac{ue} is moving, the VA remains static.

\subsection{Measurement Models} \label{Measurement model}
Every time step $k$, the \ac{bs} sends downlink signals, which reach the \ac{ue} via the line-of-sight (LOS) path as well as non-line-of-sight (NLOS) paths, via SPs or reflecting surfaces. We consider OFDM transmissions, thus the received signal at subcarrier $s$ and time step $k$ can be expressed as\cite{heath2016overview}
\begin{align}
    &\boldsymbol{Y}_{s,k}=\boldsymbol{W}_{k}^{\mathsf{H}}\sum _{i=0}^{I_{k}-1}g_{k}^{i}\boldsymbol{a}_{\text{R}}(\boldsymbol{\theta}_{k}^{i})\boldsymbol{a}_{\text{T}}^{\mathsf{H}}(\boldsymbol{\phi}_{k}^{i})e^{-\jmath 2\pi s \Delta f \tau_{k}^{i}}\boldsymbol{S}_s+ \boldsymbol{N}_{s,k}, %\notag
    \label{reveivedsignal}
\end{align}
where $\boldsymbol{S}_s$ is the pilot signal over subcarrier $s$ (including possible precoding); $\boldsymbol{Y}_{s,k}$ is the received signal over subcarrier $s$; $\boldsymbol{W}_{k}$ is a combining matrix. The number of visible landmarks is denoted as $I_{k}$, while index $i=0$ corresponds to the \ac{bs}. We further assume that there is only one path from a landmark. %, so the path with index $i=0$ is the line-of-sight (LOS), and the remains are non-line-of-sights (NLOS). 
Moreover, $\boldsymbol{a}_{\text{R}}(\cdot)$ and $\boldsymbol{a}_{\text{T}}(\cdot)$ are the steering vectors of the receiver and transmitter antenna arrays, respectively; $\Delta f$ is the subcarrier spacing; $\boldsymbol{N}_{s,k}$ is the noise.
Each path $i$ can be described by a complex gain $g_{k}^{i}$, a time of arrival (TOA) $\tau_{k}^{i}$, an angle of arrival (AOA) pair $\boldsymbol{\theta}_{k}^{i}$ in azimuth and elevation, and an angle of departure (AOD) pair $\boldsymbol{\phi}_{k}^{i}$ in azimuth and elevation. The relations between the channel parameters and the sensor and landmark states can be found, e.g., in \cite[Appendix A]{ge20205GSLAM}.
These parameters are estimated by a parametric channel estimation algorithm, such as  \cite{richter2005estimation,alkhateeb2014channel,venugopal2017channel,Gershman2010,jiang2021high}. 

 %The TOA, AOA and AOD depend on the geometric relation among the transmitter, the receiver, and the incident points of NLOS paths in the environment. 
%\subsection{Measurement Model} 
%At the receiver side, a channel estimator can be executed to provide estimates of angles and delays of paths from the received signal. There are several existing methods for channel estimation\footnote{However, the channel estimation is out of the scope of this paper, and the \ac{ue} directly utilizes the output estimates of the channel parameters.}, such as search-based and search-free methods \cite{richter2005estimation,alkhateeb2014channel,venugopal2017channel,Gershman2010,jiang2021high}. 
The channel estimator provides a set $\mathcal{Z}_{k}$ at time $k$, 
with elements $\{\boldsymbol{z}_{k}^{1},\dots, \boldsymbol{z}_{k}^{\hat{{I}}_{k}} \}$. In general, $\hat{{I}}_{k} \neq {{I}}_{k}$, since measurements may originate from clutter (e.g., due to transient objects or noise peaks during channel estimation) and landmarks may be misdetected. The clutter can be modeled as a Poisson point process (PPP) with clutter intensity $c(\boldsymbol{z})$. To account for misdetections,  we  introduce the detection probability $p_{\text{D}}(\boldsymbol{x}^{i},\boldsymbol{s}_{k}) \in [0,1]$ 
that landmark $\boldsymbol{x}^{i}$ is detected with a measurement when the sensor has state $\boldsymbol{s}_{k}$.
%, denoted as $p_{\text{D}}$ as the short-hand, and the survival probability $p_{\text{S}}(\boldsymbol{x}_{k}^{i},\boldsymbol{s}_{k}) \in [0,1]$ to represent the possibility that $\boldsymbol{x}_{k}^{i}$ still exists, denoted as $p_{\text{S}}$ as the short-hand.
The measurement originating from a landmark is characterized by a likelihood function
\begin{align}
    f(\boldsymbol{z}_{k}^{i}|\boldsymbol{x}^{i},\boldsymbol{s}_{k})=\mathcal{N}(\boldsymbol{z}_{k}^{i};\boldsymbol{h}(\boldsymbol{x}^{i},\boldsymbol{s}_{k}),\mathbf{R}_k^i),\label{pos_to_channelestimation}
\end{align}
%
%Each measurement follows
%\begin{equation}
 %   \boldsymbol{z}_{k}^{i}=\boldsymbol{h}(\boldsymbol{x}_{k}^{i},\boldsymbol{s}_{k})+\boldsymbol{w}_{k}^{i}, \label{pos_to_channelestimation}
%\end{equation}
where %$\boldsymbol{x}^{i}$ denotes the associated landmarks state to the measurement $\boldsymbol{z}_{k}^{i}$, 
$\boldsymbol{h}(\boldsymbol{x}^{i},\boldsymbol{s}_{k})=[\tau_{k}^{i},(\boldsymbol{\theta}_{k}^{i})^{\mathsf{T}},(\boldsymbol{\phi}_{k}^{i})^{\mathsf{T}}]^{\mathsf{T}}$ %is the function that transforms the geometric information to the TOA, AOAs and AODs, 
and $\mathbf{R}_k^i$ is the measurement covariance. We also introduce the state variable $\tilde{\boldsymbol{s}}^{i}_k=[\boldsymbol{s}^{\text{T}}_k,\boldsymbol{x}^{\text{T}}_i]^{\text{T}}$, allowing us to write $\boldsymbol{h}(\tilde{\boldsymbol{s}}^{i}_k)$.

%, and the noise covariance $\mathbf{R}_k^i$ can be computed. The number of measurements is denoted as $\hat{\boldsymbol{I}}_{k}$, which could be different from $\boldsymbol{I}_{k}$, as the landmark could be mis-detected, due to the limitation of the receiver and the channel estimator, or the measurement could be clutter, which is caused by noise peaks. The clutter can be modeled as a Poisson point process (PPP) with clutter intensity $c(\boldsymbol{z})$. In addition, 

%\begin{equation}
%\begin{aligned}
 %   \boldsymbol{h}(\boldsymbol{x}_{k}^{i,l},\boldsymbol{s}_{k})=\left[\begin{array}{c}\tau_{k}^{i,l},(\boldsymbol{\theta}_{k}^{i,l})^{\mathsf{T}},(\boldsymbol{\phi}_{k}^{i,l})^{\mathsf{T}}\end{array}\right]^{\mathsf{T}},\label{h&noise}
%\end{aligned}
%\end{equation}

%\vspace{-1mm}

\section{Basics of the PMBM SLAM Filter} \label{PMBM_SLAM_Filter}
In this section, we briefly describe basics of the PMBM and PMB SLAM filters, which will be the benchmark and starting point for the EK-PMB(M) filters in the subsequent sections. 

\subsection{PMBM and PMB Densities}
%\Hyowon{If you want to shrink the paper, you could remove those two sentences related to  PMBM conjugacy.} 
%\textcolor{gray}{A PMBM density is a multi-object conjugate prior \cite{williams2015marginal}. Therefore, given a prior of the PMBM form, all the subsequent predicted and posterior densities are PMBM.} 
An RFS is denoted by $\mathcal{X}=\{\boldsymbol{x}_1,\ldots,\boldsymbol{x}_n\}$, where the element  $\boldsymbol{x}_i$ (indexed by $i$) denotes a random vector, and $\lvert \mathcal{X} \rvert = n$ is the random cardinality, with a set density $f(\mathcal{X})$.
% An RFS variable $\mathcal{X}$ is a random variable that takes values as
% unordered finite sets. An RFS $\mathcal{X}$ is completely specified by its density $f(\mathcal{X})$, which can be evaluated for any  $\mathcal{X}=\{\boldsymbol{x}_1,\ldots,\boldsymbol{x}_n\}$, including $\mathcal{X}=\emptyset$.
% A PMBM RFS $\mathcal{X}$ is a particular RFS random variable, which is the union of two disjoint RFSs, a PPP RFS and a MBM RFS~\cite{williams2015marginal,garcia2018poisson,fatemi2017poisson}.
When $\mathcal{X}_\mathrm{U}$ and $\mathcal{X}_\mathrm{D}$ are two independent RFSs, following  a PPP and a MBM processes, respectively, we can say that $\mathcal{X}=\mathcal{X}_\mathrm{U} \uplus \mathcal{X}_\mathrm{D}$ follows a PMBM density~\cite{williams2015marginal,garcia2018poisson,fatemi2017poisson}.
The PPP RFS $\mathcal{X}_{\mathrm{U}}$ is used to model the set of undetected landmarks, which have never been detected before.
The MBM RFS $\mathcal{X}_\text{D}$ is used to model the set of detected landmarks, which have been detected at least once before.
Then, the density of PMBM RFS is given by
\begin{align}
    f_{\mathrm{PMBM}}(\mathcal{X})=\sum_{\mathcal{X}_{\mathrm{U}}\uplus\mathcal{X}_{\mathrm{D}}=\mathcal{X}}f_{\mathrm{PPP}}(\mathcal{X}_{\mathrm{U}})f_{\mathrm{MBM}}(\mathcal{X}_{\mathrm{D}}),\label{PMBM}
\end{align}
%the set of undetected objects $\mathcal{X}_{\mathrm{U}}$ and the set of detected objects $\mathcal{X}_{\mathrm{D}}$\cite{garcia2018poisson}. The undetected objects are the objects that have never been detected before; the detected objects are the objects that have been detected at least once before. We model $\mathcal{X}_{\mathrm{U}}$ as a Poisson point process (PPP), $\mathcal{X}_{\mathrm{D}}$ as a multi-Bernoulli mixture (MBM). The details of the densities of PPP and MBM can be found in \cite{williams2015marginal,garcia2018poisson,fatemi2017poisson}. 
% Then, $f_{\mathrm{PMBM}}(\mathcal{X})$ can be defined by \cite{mahler2014advances} %the FISST convolution as 
% \begin{equation}
%     f_{\mathrm{PMBM}}(\mathcal{X})=\sum_{\mathcal{X}_{\mathrm{U}}\uplus\mathcal{X}_{\mathrm{D}}=\mathcal{X}}f_{\mathrm{PPP}}(\mathcal{X}_{\mathrm{U}})f_{\mathrm{MBM}}(\mathcal{X}_{\mathrm{D}}),\label{PMBM}
% \end{equation}
where $f_{\mathrm{PPP}}(\cdot)$ is a PPP density and $f_{\mathrm{MBM}}(\cdot)$ is an MBM density\footnote{To understand the $\uplus$ notation, consider the example where $\mathcal{X}=\{\boldsymbol{x}_1,\boldsymbol{x}_2\}$. Then the summation in \eqref{PMBM} has four terms: (i) $\mathcal{X}_{\mathrm{U}}=\emptyset$ and $\mathcal{X}_{\mathrm{D}}=\{\boldsymbol{x}_1,\boldsymbol{x}_2\}$; (ii) $\mathcal{X}_{\mathrm{U}}=\{\boldsymbol{x}_1\}$ and $\mathcal{X}_{\mathrm{D}}=\{\boldsymbol{x}_2\}$; (iii) $\mathcal{X}_{\mathrm{U}}=\{\boldsymbol{x}_2\}$ and $\mathcal{X}_{\mathrm{D}}=\{\boldsymbol{x}_1\}$; and (iv)  $\mathcal{X}_{\mathrm{U}}=\{\boldsymbol{x}_1,\boldsymbol{x}_2\}$ and $\mathcal{X}_{\mathrm{D}}=\emptyset$.}. The PPP density is given by
\begin{equation}
    f_{\mathrm{PPP}}(\mathcal{X}_{\mathrm{U}})=e^{-\int\mathrm{D}_{\mathrm{u}}(\boldsymbol{x}')\mathrm{d}\boldsymbol{x}'} \prod_{\boldsymbol{x} \in \mathcal{X}_{\mathrm{U}}} \mathrm{D}_{\mathrm{u}}(\boldsymbol{x}) %\mathcal{X}_{\mathrm{U}}}  \mathrm{D}_{\mathrm{u}}^{\mathcal{X}_{\mathrm{U}}}
    ,\label{PPP}
\end{equation}
where $\mathrm{D}_{\mathrm{u}}(\cdot)$ is the intensity function,
% for $\mathcal{X}_{\mathrm{U}}$, 
%Hence, \eqref{PPP} can be parameterized by $\mathrm{D}_{\mathrm{u}}(\boldsymbol{x})$ \Hyowon{this sentence could be moved to below}. 
% while 
and the MBM density follows
\begin{equation}
    f_{\mathrm{MBM}}(\mathcal{X}_{\mathrm{D}})= \sum_{j\in\mathbb{I}}w^{j}\sum_{\uplus_{i \in \mathbb{I}^{j}}  \mathcal{X}^{i}=\mathcal{X}_{\mathrm{D}}}\prod_{i=1}^{n}f^{j,i}_{\mathrm{B}}(\mathcal{X}^{i}),\label{MBM}
\end{equation}
where $\mathbb{I}$ is the index  set of the global hypotheses~\cite{williams2015marginal}; $w^{j}$ is the weight for global hypothesis $j$, satisfying $\sum_{j\in\mathbb{I}}w^{j}=1$; $n$ is the number of potentially detected landmarks (for convenience set to the same value for all global hypotheses); $\mathbb{I}^{j}$ is the index set of landmarks (i.e., the Bernoulli components) under global hypothesis $j$; and $f_{\mathrm{B}}^{j,i}(\cdot)$ is the Bernoulli density of landmark $i$ under global hypothesis $j$. Each Bernoulli density follows
\begin{equation}
f^{j,i}_{\mathrm{B}}(\mathcal{X}^{i})=
\begin{cases}
1-r^{j,i} \quad& \mathcal{X}^{i}=\emptyset, \\ r^{j,i}f^{j,i}(\boldsymbol{x}) \quad & \mathcal{X}^{i}=\{\boldsymbol{x}\}, \\ 0 \quad & \mathrm{otherwise},
\end{cases}
\end{equation} 
%and $f^{j,i}_{\mathrm{B}}(\mathcal{X}^{j})=0$  otherwise. 
where $r^{j,i}\in [0,1]$ is the existence probability of the landmark and $f^{j,i}(\boldsymbol{x})$ is the pdf of the vector $\boldsymbol{x}$.
% state density. 
%\Hyowon{I guess you could remove the sentences from here.}Hence, \eqref{MBM} can be parameterized by $\{l^{j,i},\{r^{j,i},f^{j,i}(\boldsymbol{x})\}_{j\in \mathbb{I}^{h}}\}_{h\in \mathbb{I}}$, where $\mathbb{I}$ is the index set.Hence, the PMBM  is fully described  $\mathrm{D}_{\mathrm{u}}(\boldsymbol{x})$ and $\{l^{j,i},\{r^{j,i},f^{j,i}(\boldsymbol{x})\}_{j\in \mathbb{I}^{h}}\}_{h\in \mathbb{I}}$\cite{williams2015marginal}.
\begin{example}
Consider a PMBM with $|\mathbb{I}|=2$ components, each MB contains a single Bernoulli (i.e., $|\mathbb{I}^1|=|\mathbb{I}^2|=n=1$). Suppose we evaluate the PMBM in $\mathcal{X}=\{\boldsymbol{x}\}$ (i.e., containing a single element). Then, 
\begin{align*}
    f_{\mathrm{PMBM}}(\{\boldsymbol{x}\})  =& f_{\mathrm{P}}(\{\boldsymbol{x}\})(w^1(1-r^{1,1})+w^2(1-r^{2,1}))+ f_{\mathrm{P}}(\emptyset)(w^1r^{1,1}f^{1,1}(\boldsymbol{x})+w^2r^{2,1}f^{2,1}(\boldsymbol{x}))
\end{align*}
where $w^2 = 1-w^1$. Similarly, for $\mathcal{X}=\emptyset$ (i.e., containing no elements), we find that 
\begin{align*}
    f_{\mathrm{PMBM}}(\emptyset) & = f_{\mathrm{P}}(\emptyset)(w^1(1-r^{1,1})+w^2(1-r^{2,1})).
\end{align*}
\end{example}

In conclusion, a PMBM is described by a PPP from \eqref{PPP} and a MBM from \eqref{MBM}, which is parameterized by $\mathrm{D}_{\mathrm{u}}(\boldsymbol{x})$ and $\{w^{j},\{r^{j,i},f^{j,i}(\boldsymbol{x})\}_{i\in \mathbb{I}^{j}}\}_{j\in \mathbb{I}}$. If there is only one mixture component in the MBM (i.e., $|\mathbb{I}|=1$), then \eqref{PMBM} reduces to a PMB.

\subsection{Bayesian Recursion of RFS SLAM}\label{sec:BayesianSLAM}
An RFS SLAM filter follows the prediction and update steps of the Bayesian filtering recursion with RFSs.
% , using the Chapman-Kolmogorov equation applied to sets \cite{mahler2003multitarget}.
We denote a sensor trajectory at time $k$ by $\boldsymbol{s}_{0:k}$ and denote a set of landmarks by $\mathcal{X}$. The joint posterior density of sensor trajectory and set of landmarks can be factorized as
% for the sensor trajectory $\boldsymbol{s}_{0:k}$ and the map $\mathcal{X}$ at time step $k$, denoted as $f(\boldsymbol{s}_{0:k},\mathcal{X}|\mathcal{Z}_{1:k})$, can be factorized as 
\begin{align}
    &f(\boldsymbol{s}_{0:k},\mathcal{X}|\mathcal{Z}_{1:k}) = f(\boldsymbol{s}_{0:k}|\mathcal{Z}_{1:k}) f(\mathcal{X}|\boldsymbol{s}_{0:k},\mathcal{Z}_{1:k}),\label{prior_fact}
\end{align}
where $f(\boldsymbol{s}_{0:k}|\mathcal{Z}_{1:k})$ is the density of sensor trajectory, and $f(\mathcal{X}|\boldsymbol{s}_{0:k},\mathcal{Z}_{1:k})$ is the set density of  landmarks conditioned on the sensor trajectory.
% are the densities the sensor trajectory and the landmark conditioned on the sensor trajectory, respectively. 
%$f(\boldsymbol{s}_{0:k}|\mathcal{Z}_{1:k})$, $f(\mathcal{X}|\boldsymbol{s}_{0:k},\mathcal{Z}_{1:k})$, the transition density of a known dynamic model for the \ac{ue} state $f(\boldsymbol{s}_{k+1}|\boldsymbol{s}_{k})$, and the RFS-likelihood of the received measurement set  $g(\mathcal{Z}_{k+1}|\mathcal{Z}_{1:k},\boldsymbol{s}_{0:k+1},\mathcal{X})$,

Using the Chapman-Kolmogorov equation, the sensor prediction step is given by 
%and the update steps are given by
%\begin{itemize}
%\item[(a)] prediction: The \ac{ue} state prediction can be
\begin{align}
    f(\boldsymbol{s}_{0:k+1}|\mathcal{Z}_{1:k}) =  f(\boldsymbol{s}_{0:k}|\mathcal{Z}_{1:k})f(\boldsymbol{s}_{k+1}|\boldsymbol{s}_{k}),\label{predicted_prior_vehicle}
\end{align}
% where $f(\boldsymbol{s}_{k+1}|\mathcal{Z}_{1:k})$ is the predicted prior of the sensor state. 
where $f(\boldsymbol{s}_{k+1}|\boldsymbol{s}_{k})$ is the transition density of the dynamics in~\eqref{dynamicmodel}.
We assume landmarks are static and never appear or disappear, and thus there is no prediction for landmarks. Then, $f(\mathcal{X}|\boldsymbol{s}_{0:k+1},\mathcal{Z}_{1:k}) = f(\mathcal{X}|\boldsymbol{s}_{0:k},\mathcal{Z}_{1:k})$.
% $\mathcal{X}$. 
% Thus, the predicted density of $\mathcal{X}$ is $f(\mathcal{X}|\boldsymbol{s}_{0:k+1},\mathcal{Z}_{1:k}) = f(\mathcal{X}|\boldsymbol{s}_{0:k},\mathcal{Z}_{1:k})$. 
%\begin{align}
 %   f(\mathcal{X}|\boldsymbol{s}_{0:k+1},\mathcal{Z}_{1:k}) = f(\mathcal{X}|\boldsymbol{s}_{0:k},\mathcal{Z}_{1:k}).\label{predicted_prior_map}
%\end{align}

The joint posterior is updated as
% then  
%\item[(b)] update: The joint posterior density can be denoted by
\begin{align}
    & f(\boldsymbol{s}_{0:k+1},\mathcal{X}|\mathcal{Z}_{1:k+1})  = \label{jointposter2} \frac{g(\mathcal{Z}_{k+1}|\boldsymbol{s}_{k+1},\mathcal{X})f(\boldsymbol{s}_{0:k+1}|\mathcal{Z}_{1:k})  f(\mathcal{X}|\boldsymbol{s}_{0:k+1},\mathcal{Z}_{1:k})}{f(\mathcal{Z}_{k+1}|\mathcal{Z}_{1:k})}  %\notag 
\end{align}
where $f(\mathcal{Z}_{k+1}|\mathcal{Z}_{1:k})$ is the normalizing factor, and $g(\mathcal{Z}_{k+1}|\boldsymbol{s}_{k+1},\mathcal{X})$ is the RFS likelihood function, given by  \cite[eqs.\,(5)--(6)]{garcia2018poisson}
\begin{align}
    & g(\mathcal{Z}_{k+1}|\boldsymbol{s}_{k+1},\{\boldsymbol{x}_1,\ldots, \boldsymbol{x}_n \}) =e^{-\int c(\boldsymbol{z}) \mathrm{d} \boldsymbol{z}} \sum_{\mathcal{Z}^c\uplus\mathcal{Z}^1 \ldots \uplus \mathcal{Z}^n=\mathcal{Z}_{k+1}}\prod_{\boldsymbol{z} \in \mathcal{Z}^c}c(\boldsymbol{z})\prod_{l=1}^{n}\ell(\mathcal{Z}^i|\boldsymbol{s}_{k+1},\boldsymbol{x}_i),
\end{align}
where $\ell(\cdot)$ follows
\begin{align}
    \ell(\mathcal{Z}^i |\boldsymbol{s}_{k+1},\boldsymbol{x}_i)=
\begin{cases}
 1-p_{\text{D}}(\boldsymbol{x}^{i},\boldsymbol{s}_{k+1}) \quad & \mathcal{Z}^{i}=\emptyset, \\ p_{\text{D}}(\boldsymbol{x}^{i},\boldsymbol{s}_{k+1})f(\boldsymbol{z}|\boldsymbol{x}^{i},\boldsymbol{s}_{k+1}) \quad& \mathcal{Z}^{i}=\{\boldsymbol{z} \}, \\0 \quad & \mathrm{otherwise}.
\end{cases}
\end{align}
%is the non-zero for $\ell(\{\boldsymbol{z} \}|\boldsymbol{s}_{k+1},\boldsymbol{x}_i)=p_{\text{D}}(\boldsymbol{x}^{i},\boldsymbol{s}_{k+1})f(\boldsymbol{z}|\boldsymbol{x}^{i},\boldsymbol{s}_{k+1})$ and $\ell(\emptyset |\boldsymbol{s}_{k+1},\boldsymbol{x}_i)=1-p_{\text{D}}(\boldsymbol{x}^{i},\boldsymbol{s}_{k+1})$.

\subsection{Overview of PMBM and PMB SLAM Filters}
%The PMBM filter follows the prediction and update steps of the Bayesian filtering recursion with RFSs, using the Chapman-Komogorov equation applied to sets \cite{mahler2003multitarget}. 
If the PMBM density is the conjugate prior for the transition density and the measurement model, then all subsequent predicted and updated distributions by the Bayesian recursion (see, Section~\ref{sec:BayesianSLAM}) will preserve the PMBM form with parameters~\cite{williams2015marginal}:
% of the density \cite{vo2013labeled}. 
% The PMBM parameters
$\mathrm{D}_{\mathrm{u}}(\boldsymbol{x})$ and $\{w^{j},\{r^{j,i},f^{j,i}(\boldsymbol{x})\}_{i\in \mathbb{I}^{j}}\}_{j\in \mathbb{I}}$.
% go through the Bayesian filtering recursion comprising of prediction and update steps.
% filtering recursions can then be expressed in terms of prediction and update steps for the PMBM parameters  $\mathrm{D}_{\mathrm{u}}(\boldsymbol{x})$ and $\{w^{j},\{r^{j,i},f^{j,i}(\boldsymbol{x})\}_{i\in \mathbb{I}^{j}}\}_{j\in \mathbb{I}}$.
% The PMBM SLAM filter relies on a PMBM density representation for the landmarks, \emph{conditioned} on the sensor state trajectory.
% In the PMBM SLAM density, the set density of
The landmark is \emph{conditioned} on the sensor state trajectory, and the set densities $f(\mathcal{X}|\boldsymbol{s}_{0:k},\mathcal{Z}_{1:k})$ and $f(\mathcal{X}|\boldsymbol{s}_{0:k},\mathcal{Z}_{1:k-1})$ are PMBM.

The PMBM SLAM filter can be implemented using a \ac{rbpf}, where the sensor trajectory density is represented by particle samples~\cite{ge20205GSLAM}: $f(\boldsymbol{s}_{0:k+1}|\mathcal{Z}_{1:k+1}) \approx \sum_n^N \boldsymbol{s}_{0:k+1}^{(n)}w_{k+1}^{(n)}$, where $n$ is the particle index; $N$ is the number of particle samples; $\boldsymbol{s}_{0:k+1}^{(n)}$ is the particle sample; and $w_{k+1}^{(n)}$ is the particle weight such that $\sum_n w_{k+1}^{(n)}=1$.
% $\boldsymbol{s}^{(n)}_{0:k}$, $n=1,\ldots, N$ \cite{ge20205GSLAM}, where $N$ is the number of particles.
The prediction step \eqref{predicted_prior_vehicle} then follows the standard particle generation of $\boldsymbol{s}^{(n)}_{0:k+1}$, while the update step \eqref{jointposter2} requires  computation of  $f(\mathcal{X}|\boldsymbol{s}^{(n)}_{0:k+1},\mathcal{Z}_{1:k+1})$ and $w_{k+1}^{(n)}$ for each particle.

In the PMB SLAM filter, under a \ac{rbpf} implementation,  $f(\mathcal{X}|\boldsymbol{s}^{(n)}_{0:k},\mathcal{Z}_{1:k})$ is given by a PMB density.
After the update step, $f(\mathcal{X}|\boldsymbol{s}^{(n)}_{0:k+1},\mathcal{Z}_{1:k+1})$ is possibly a PMBM
since the filter tracks possible association hypotheses to the measurements.
% , due to new global hypotheses that are generated
% The TOMB/P algorithm can then be applied to approximate the PMBM to a PMB for the next time step. 
We adopt the method of marginal association distribution~\cite{williams2015marginal}, which enables us to approximate the PMBM to a PMB at the end of time step by marginalizing over the data association.

%In the PMBM SLAM filter, the large number of MBs makes the PMBM SLAM filter intractable, and approximations must be made. One solution is to approximate the discrete distribution of data association and 
%By marginalizing over the data associations the the PMB SLAM filter. 

\subsubsection*{Complexity}
% In the PMBM SLAM filter, at every time step, each possible data association under a previous global hypothesis can become a new global hypothesis.
\sloppy At the end of every time step, each previous global hypothesis considers all possible data associations, which generates a variety of new global hypotheses, and the number of global hypotheses rapidly increases in combinatorial explosion.
% , which makes the number of MBs increases combinatorially. 
In particular, the number of global hypotheses per each particle at time step $k+1$ is given by $|\mathbb{I}_{k+1}|=\sum_{j \in \mathbb{I}_{k}}\sum_{\alpha=0}^{|\mathcal{Z}_{k+1}|}\mathrm{C}_{|\mathcal{Z}_{k+1}|}^{\alpha}\mathrm{A}_{|\mathbb{I}^{j}_{k}|}^{|\mathcal{Z}_{k+1}|-\alpha}$, which grows over time
% rapidly in $k$  
\cite[Appendix B]{ge20205GSLAM}. Here, $\mathrm{C}$ and  $\mathrm{A}$ denote the  combination and permutation operations, respectively. The complexity,  at time step $k+1$, scales  as $\mathcal{O}(N\sum_{j \in \mathbb{I}_{k}}|\mathbb{I}^{j}_{k}|\sum_{\alpha=0}^{|\mathcal{Z}_{k+1}|}\mathrm{C}_{|\mathcal{Z}_{k+1}|}^{\alpha}\mathrm{A}_{|\mathbb{I}^{j}_{k}|}^{|\mathcal{Z}_{k+1}|-\alpha})$, where number of particles $N$ can be in the order of $1000-10000$, depending on the state dimension.

In the PMB filter, the number of global hypotheses,
% MBs, 
at time step $k+1$, is $\sum_{\alpha=0}^{|\mathcal{Z}_{k+1}|}\mathrm{C}_{|\mathcal{Z}_{k+1}|}^{\alpha}\mathrm{A}_{|\mathbb{I}^{1}_{k}|}^{|\mathcal{Z}_{k+1}|-\alpha}$, which is much smaller than the number of global hypotheses
% MBs 
in the PMBM SLAM filter. Then, the marginal association distribution
% TOMB/P algorithm
is used to keep $f(\mathcal{X}|\boldsymbol{s}^{(n)}_{0:k+1},\mathcal{Z}_{1:k+1})$ as a PMB density. The complexity, at time step $k$, scales  as $\mathcal{O}(N|\mathbb{I}^{1}_{k}|\sum_{\alpha=0}^{|\mathcal{Z}_{k+1}|}\mathrm{C}_{|\mathcal{Z}_{k+1}|}^{\alpha}\mathrm{A}_{|\mathbb{I}^{1}_{k}|}^{|\mathcal{Z}_{k+1}|-\alpha})$.

In summary, both the PMB and PMBM suffer from high complexity, due to both the large number of hypotheses and the number of particles in the \ac{rbpf} implementation.

%
%Given this joint prior, the transition density of a known dynamic model for the \ac{ue} state $f(\boldsymbol{s}_{k+1}|\boldsymbol{s}_{k})$, and the likelihood  of the received measurement set  $g(\mathcal{Z}_{k+1}|\mathcal{Z}_{1:k},\boldsymbol{s}_{k+1},\mathcal{X})$, the joint posterior density can be written as

%$f(\mathcal{X}|\boldsymbol{s}_{k+1},\mathcal{Z}_{1:k})$ is the predicted prior of the map, denoted as 
%\begin{align}
%    f(\mathcal{X}|\boldsymbol{s}_{0:k+1},\mathcal{Z}_{1:k}) = \frac{f(\mathcal{X}|\boldsymbol{s}_{0:k},\mathcal{Z}_{1:k})f(\boldsymbol{s}_{k+1}|\boldsymbol{s}_{k})}{ f(\boldsymbol{s}_{0:k+1}|\mathcal{Z}_{1:k})},\label{predicted_prior_map}
%\end{align}

%\subsection{PMB SLAM} \label{sec:PMBSLAM}
%In the PMBM SLAM filter, large number of MBs makes the PMBM SLAM filter intractable, and approximations mush be made. One solution is to approximate the discrete distribution of data association and marginalize over it, which can be done by using the so called \ac{tomb} algorithm \cite{williams2015marginal}, resulting in the PMB SLAM filter. 

\section{Proposed EK-PMBM SLAM Filter} \label{EKF-PMBM}
%\subsection{Principles}

To reduce the computational cost, we motivate the joint vehicle and landmark update using the EK filter~\cite[Ch. 5.2]{sarkka2013bayesian} instead of the RBPF.
We will introduce the marginal posterior densities for SLAM and their Bayesian recursion.
Then, we describe a novel algorithm of PMBM SLAM filter with the joint update step and its EK filter implementation.

\subsection{Form of the EK-PMBM Filter}
% The EKF-PMBM relies on the EKF \cite[Ch. 5.2]{sarkka2013bayesian} and thus 
\sloppy We suppose that at time step $k$, the vehicle state density is a Gaussian distribution ${\cal N}(
% \boldsymbol{s}_{k|k}
\boldsymbol{s}_k
; \boldsymbol{m}_{k|k},\boldsymbol{P}_{k|k})$, where $\boldsymbol{m}_{k|k}$ and $\boldsymbol{P}_{k|k}$ are the mean and the covariance matrix, respectively; the PPP parameter  $\lambda_{k|k}(\boldsymbol{x})$, which can be modeled as a $\eta_{k|k}\mathfrak{U}(\boldsymbol{x}) $ with $\mathfrak{U}(\boldsymbol{x}) $ representing a uniform distribution over the space, and MBM parameters $\{w^{j}_{k},\{r_{k|k}^{j,i},f_{k|k}^{j,i}(\boldsymbol{x})\}_{i\in \mathbb{I}_{k}^{j}}\}_{j\in \mathbb{I}_{k}}$ for the map are also given, where each $f_{k|k}^{j,i}(\boldsymbol{x})$ is a Gaussian distribution ${\cal N}(\boldsymbol{x}^{j,i}; \boldsymbol{u}_{k|k}^{j,i},\boldsymbol{C}^{j,i}_{k|k})$.  %where $\boldsymbol{u}_{k|k}$ and $\boldsymbol{C}^{j,i}_{k|k}$ are the mean and the covariance matrix, respectively. 
Therefore, the MBM parameters can be written as  $\{w^{j}_{k},\{r_{k|k}^{j,i},\boldsymbol{u}_{k|k}^{j,i},\boldsymbol{C}^{j,i}_{k|k} \}_{i\in \mathbb{I}_{k}^{j}}\}_{j\in \mathbb{I}_{k}}$.

% The proposed filter aims to address the two aforementioned computational bottlenecks, and is based on three ideas: i)  state marginalization per time step, ii) separate sensor state and map prediction using the EKF, iii) joint sensor state and map update using the EKF. 

%To efficiently compute \eqref{eq:UpMarObjMarg1} and \eqref{eq:UpMarObjMarg2}, we adopt the EKF to handle the non-linearity, which forms the first order Taylor series-based Gaussian approximation to the filtering distribution \cite[Ch. 5.2]{sarkka2013bayesian}. 

\subsection{Marginalization of PMBM SLAM Density}
At each time step, rather than keeping 
% track of 
the entire state trajectory, we keep track of marginal posteriors $f(\mathcal{X}|\mathcal{Z}_{1:k})$ and $f(\mathbf{s}_{k}|\mathcal{Z}_{1:k})$ \cite{kim2020low,frohle2019multisensor}. This implies that this posterior no longer carries the correlation between the sensor state trajectory and the map state, which constitutes an inherent loss of information and is the price to pay for reducing complexity. % so that in turn we no longer need to keep track of the entire vehicle trajectory. 
The prediction step then simplifies to 
\begin{align}
     f(\boldsymbol{s}_{k+1}|\mathcal{Z}_{1:k}) & = \int f(\boldsymbol{s}_{k}|\mathcal{Z}_{1:k})f(\boldsymbol{s}_{k+1}|\boldsymbol{s}_{k}) \text{d} \boldsymbol{s}_{k}.\label{predicted_prior_vehicleMarg}
\end{align} 
The update step for the sensor state becomes
\begin{align}
   &  f(\mathbf{s}_{k+1}|\mathcal{Z}_{1:k+1})  = \int  f(\boldsymbol{s}_{k+1},\mathcal{X}|\mathcal{Z}_{1:k+1})\delta \mathcal{X}  \label{eq:UpMarObjMarg1}\\
    %& = \frac{1}{f(\mathcal{Z}_{k+1}|\mathcal{Z}_{1:k})}
    & \propto \int f(\mathcal{X}|\mathcal{Z}_{1:k})f(\mathbf{s}_{k+1}|\mathcal{Z}_{1:k}) g(\mathcal{Z}_{k+1}|\mathbf{s}_{k+1},\mathcal{X}) \delta \mathcal{X},
\end{align}
whereas for the map state we find that 
\begin{align}
    &f(\mathcal{X}|\mathcal{Z}_{1:k+1}) = \int f(\boldsymbol{s}_{k+1},\mathcal{X}|\mathcal{Z}_{1:k+1}) \mathrm{d}\mathbf{s}_{k+1}\label{eq:UpMarObjMarg2}\\
    %& = \frac{1}{f(\mathcal{Z}_{k+1}|\mathcal{Z}_{1:k})} 
    & \propto \int f(\mathcal{X}|\mathcal{Z}_{1:k})f(\mathbf{s}_{k+1}|\mathcal{Z}_{1:k}) g(\mathcal{Z}_{k+1}|\mathbf{s}_{k+1},\mathcal{X})
    \mathrm{d}\mathbf{s}_{k+1}.
\end{align}
In \eqref{eq:UpMarObjMarg1}, $\int g'(\mathcal{X})\delta \mathcal{X}$ refers to the set integral \cite[eq.~(4)]{williams2015marginal}.

\subsection{EK-PMBM SLAM Prediction} \label{EKF-pred}
%The PMBM SLAM filter goes through the prediction steps \eqref{predicted_prior_vehicle}, \eqref{predicted_prior_map}, and joint update step \eqref{jointposter} recursively. 

%\subsubsection{Initial Condition at Time $k$}

%\subsubsection{Prediction}
Following \eqref{predicted_prior_vehicleMarg}, the predicted \ac{ue} state at the time step $k+1$ can be acquired via first-order EK filter \cite[Ch. 5.2]{sarkka2013bayesian} 
\begin{align}
    &\boldsymbol{m}_{k+1|k} = \boldsymbol{v}(\boldsymbol{m}_{k|k})\label{EKF_predicted_mean_vehicle}\\&
    \boldsymbol{P}_{k+1|k} = \boldsymbol{F}_{k|k} \boldsymbol{P}_{k|k} \boldsymbol{F}_{k|k}^{\text{T}}+ \boldsymbol{Q}_{k+1} ,\label{EKF_predicted_covariance_vehicle}
\end{align}
where $\boldsymbol{F}_{k|k}$ represents the Jacobian
\begin{align}
    \boldsymbol{F}_{k|k} = \left.\frac{\partial \boldsymbol{v}(\boldsymbol{s}_{k})}{\partial \boldsymbol{s}_{k}}\right|_{\boldsymbol{s}_{k}=\boldsymbol{m}_{k|k}}.
\end{align}
%of $\boldsymbol{v}(\cdot)$ with respect to $\boldsymbol{s}_{k|k}$, evaluated at $\boldsymbol{s}_{k|k}=\boldsymbol{m}_{k|k}$,  
%with elements
%\begin{align}
 %   &[\boldsymbol{F}_{k|k}]_{\alpha,\beta} = \left. \frac{\partial[\boldsymbol{v}(\boldsymbol{s}_{k|k})]_{\alpha}}{\partial [\boldsymbol{s}_{k|k}]_{\beta}} \right|_{\boldsymbol{s}_{k|k}=\boldsymbol{m}_{k|k}}.
%\end{align}

In terms of the map, 
%\subsection{Map Prediction}
since the targets are assumed to be static, there is no prediction for the landmark states and covariances. Thus, we have $\eta_{k+1|k}=\eta_{k|k}$,  
%$l_{k+1|k}^{j,i}= l_{k|k}^{j,i}$, 
$r_{k+1|k}^{j,i}=r_{k|k}^{j,i}$, $\boldsymbol{u}_{k+1|k}^{j,i}=\boldsymbol{u}_{k|k}^{j,i}$, $\boldsymbol{C}_{k+1|k}^{j,i}=\boldsymbol{C}_{k|k}^{j,i}$.

%However, as the user will enter into a new area, there will be some new landmarks appears in the field of view (FOV) and some landmarks die out. Therefore, we add a new birth intensity $\lambda_{\mathrm{B},k+1}(\boldsymbol{x})$ which represents the new landmarks entering into the FOV to the PPP, and decrease the previous PPP intensity and Bernoulli existence probabilities according the survival probability $p_{\mathrm{S}}$, which represents landmarks may die out, written as  $\lambda_{k+1|k}(\boldsymbol{x})=p_\text{S}\lambda_{k|k}(\boldsymbol{x})+\lambda_{\mathrm{B},k+1}(\boldsymbol{x})$,  $l_{k+1|k}^{j,i}= l_{k|k}^{j,i}$, $r_{k+1|k}^{j,i}=p_{\mathrm{S}}r_{k|k}^{j,i}$, $\boldsymbol{u}_{k+1|k}^{j,i}=\boldsymbol{u}_{k|k}^{j,i}$, $\boldsymbol{C}_{k+1|k}^{j,i}=\boldsymbol{C}_{k|k}^{j,i}$.

\subsection{EK-PMBM SLAM Update}  \label{joint_user_map_update}
The update is more involved and comprises the following steps. First, the data association cost metrics are computed  per MBM mixture component $j$ (i.e., per global hypothesis). Second, the top $\gamma \ge 1$ best data associations per global hypothesis are determined from the cost metrics, followed by combining of all best data associations across the global hypotheses. Third, the EK filter joint update is performed for each of the best data associations. Finally, the vehicle state density is computed, marginalizing out the best data associations. These four steps are now explained in detail.

\subsubsection{Computation of Data Association Metric} \label{Data_Association_Metric}
%In order to jointly update the user and the map states, the data association problem should be firstly solved. 
The data association cost metric depends on the local hypothesis weights, which are computed from  measurements and previously seen landmarks \cite{williams2015marginal}. We distinguish three cases\footnote{The fourth case of an undetected landmark remaining undetected does not affect the cost metric.}: 
\begin{itemize}
\item[(a)] A previously detected landmark $i$ under global hypothesis $j$ is detected again with measurement $\boldsymbol{z}^{p}_{k+1}$. The local association weight is
\begin{align}
    &l_{k+1}^{j,i,p}= r_{k+1|k}^{j,i}\rho_{k+1|k+1}^{j,i,p},\label{weight_detected}
\end{align}
where $\rho_{k+1|k+1}^{j,i,p}$ is computed as described in Appendix \ref{app:caseA}. The first factor is the existence probability of the Bernoulli, and the second factor accounts for the spatial density and the measurement likelihood.

\item[(b)] A previously detected landmark $i$ under global hypothesis $j$ is not detected at time $k+1$. The local association weight is
\begin{align}
    &l_{k+1}^{j,i,0}=(1-r_{k+1|k}^{j,i})+(1-p_{\text{D},k+1}^{j,i})r_{k+1|k}^{j,i}\label{weightmisdetected},
\end{align}
where the first term accounts for the landmark not existing in the first place, while the second term accounts for the landmark existing, but leading to a miss-detection. {Here, the detection probability is computed per landmark, as detailed in Appendix~\ref{app:caseA}.}

\item[(c)] \label{detected_first_time} A previously undetected landmark is detected for the first time with measurement $\boldsymbol{z}^{p}_{k+1}$. The local association weight is
\begin{align}
    &l_{\text{B},k+1}^{p}=c(\boldsymbol{z}^{p}_{k+1})+\rho_{\text{B},k+1|k+1}^{p}\label{weightnewdetected}
\end{align}
where $\rho_{\text{B},k+1|k+1}^{p}$ is computed,  as described in Appendix \ref{app:caseC}. The first term accounts for the fact that the measurement may be due to clutter, while the second term accounts for the fact that the measurement may be due to a new landmark from the PPP.

\end{itemize}

\subsubsection{Computation of Best $\gamma$ Data Associations} \label{Best_DA}
We construct a cost matrix $\mathbf{L}^{j}_{k+1} \in \mathbb{R}^{|\mathcal{Z}_{k+1}| \times (|\mathcal{Z}_{k+1}|+|\mathbb{I}^{j}_{k}|)}$,  using the local association  weights %calculated by \eqref{weight_detected}, \eqref{weightmisdetected} and \eqref{weightnewdetected}  for each global hypothesis
\cite{garcia2018poisson}
\begin{align}
&\mathbf{L}^{j}_{k+1} = -\ln \left[
\begin{matrix}
\tilde{l}_{k+1}^{j,1,1} & \ldots & \tilde{l}_{k+1}^{j,|\mathbb{I}^{j}_{k}|,1} \\ 
\vdots &  \ddots & \vdots \\
\tilde{l}_{k+1}^{j,1,|\mathcal{Z}_{k+1}|}  & \ldots & \tilde{l}_{k+1}^{j,|\mathbb{I}^{j}_{k}|,|\mathcal{Z}_{k+1}|}
\end{matrix}
\left|
\,
\begin{matrix}
l^{1}_{\text{B},k+1}  & \ldots & 0 \\ 
\vdots &  \ddots & \vdots \\
0 &  \ldots & l^{|\mathcal{Z}_{k+1}|}_{\text{B},k+1}
\end{matrix}
\right.
\right],%\nonumber
\end{align}
where $\tilde{l}^{j,i,p}_{k+1}=l^{j,i,p}_{k+1}/l^{j,i,0}_{k+1}$. The left $|\mathcal{Z}_{k+1}| \times |\mathbb{I}^{j}_{k}|$ sub-matrix in $\mathbf{L}^{j}_{k+1}$ corresponds to previous detections, the right $|\mathcal{Z}_{k+1}| \times |\mathcal{Z}_{k+1}|$ diagonal sub-matrix corresponds to new detections, and the off-diagonal elements of the right sub-matrix are $-\infty$. The $\gamma$-best data associations with weights representing the probability per each can be selected out by solving the assignment problem 
\begin{align}\label{optimization_problem}
\text{minimize} \quad  & \text{tr} \left(\mathbf{A}^{\text{T}} \mathbf{L}^{j}_{k+1} \right) \\
\text{s.t.} \quad  & [\mathbf{A} ]_{\alpha,\beta} \in 
\{ 0,  1 \} \quad \forall \; \alpha,\beta \nonumber \\ %, \quad \forall A^{i,j} \nonumber \\
%i,j \in \lbrace 1,\ldots,n_k \rbrace \times \lbrace 1,\ldots,n_k+m_k \rbrace \\
& \sum\nolimits_{\beta=1}^{|\mathbb{I}^{j}_{k}| + |\mathcal{Z}_{k+1}|} [ \mathbf{A} ]_{\alpha,\beta} = 1, \quad \forall \; \alpha \nonumber \\ %\quad i \in \lbrace 1,\ldots,n_k \rbrace \nonumber \\
& \sum\nolimits_{\alpha=1}^{|\mathcal{Z}_{k+1}|} [\mathbf{A} ]_{\alpha,\beta} \in  \{ 0,  1 \}, \quad \forall \; \beta \nonumber
\end{align}
using the Murty's algorithm \cite{murty1968letter}, where  $\mathbf{A}\in \mathbb{R}^{|\mathcal{Z}_{k+1}| \times (|\mathcal{Z}_{k+1}|+|\mathbb{I}^{j}_{k}|)}$ is the assignment matrix. The solutions are denoted by $\mathbf{A}^{j,h}$, where $h$ is an index
in the index set of new data associations under global hypothesis $j$, denoted as $\mathbb{H}^{j}_{k+1}$ with $|\mathbb{H}^{j}_{k+1}|\leq \gamma$. The index set of landmarks under the $j,h$-th ``new data association'' is denoted as $\mathbb{I}_{k+1}^{j,h}$, with $|\mathbb{I}_{k+1}^{j,h}|\leq |\mathbb{I}_{k}^{j}|+|\mathcal{Z}_{k+1}|$. % with the some birth components may not exist under the data association.

For the $j,h$-th data association, the corresponding assignment matrix  $\mathbf{A}^{j,h}$, can be translated to an vector, denoted as $\boldsymbol{\sigma}^{j,h}=[\sigma^{j,h}(1),\cdots,\sigma^{j,h}(|\mathbb{I}_{k}^{j}|+|\mathcal{Z}_{k+1}|)]$, defined as, for $t\leq |\mathbb{I}_{k}^{j}|$
%with each element representing the association of the $t$-th component, wh
\begin{align}
\sigma^{j,h}(t)=
\begin{cases}
p\quad& \exists ~p: [ \mathbf{A}^{j,h} ]_{p,t}=1  \\ 0 \quad & \nexists ~ p: [ \mathbf{A}^{j,h} ]_{p,t}=1,
\end{cases}
\end{align}
and for $t>|\mathbb{I}_{k}^{j}|$
\begin{align}
\sigma^{j,h}(t)=
\begin{cases}
p\quad& \exists ~p: [ \mathbf{A}^{j,h} ]_{p,t}=1  \\  \emptyset \quad & [ \mathbf{A}^{j,h} ]_{t-|\mathbb{I}_{k}^{j}|,t}=0,
\end{cases}
\end{align}
%\begin{align}
%\sigma^{j,h}(t)=
%\begin{cases}
%p\quad& \exists ~p: [ \mathbf{A}^{j,h} ]_{p,t}=1,  \\ 0 \quad & \nexists ~ p: [ \mathbf{A}^{j,h} ]_{p,t}=1, t\leq |\mathbb{I}_{k}^{j}|\\ \emptyset \quad & [ \mathbf{A}^{j,h} ]_{t-|\mathcal{Z}_{k+1}|,t-|\mathcal{Z}_{k+1}|}=0, t > |\mathbb{I}_{k}^{j}|
%\end{cases},
%\end{align}
These four cases correspond to a previously detected landmark associated to measurement $p$, a previously detected landmark being misdetected, a new landmark being associated to measurement $p$, and a new landmark being non-existent. Note that $p$ can only be $t-|\mathbb{I}_{k}^{j}|$ in the third case.
%where the three entries represent that the index of the measurement associated to the landmark, previously detected landmark is misdetected, and the newly detection is non-exist, respectively. 
%It is obvious that $\sigma^{j,h}(t) \in \{0,\cdots,|\mathcal{Z}_{k+1}|\}$ if $t \leq |\mathbb{I}_{k}^{j}|$, and $\sigma^{j,h}(t) \in \{t-|\mathcal{Z}_{k+1}|,\emptyset\}$ if $t>|\mathcal{Z}_{k+1}|$.

Each data association has a weight $w^{j,h}_{k+1}$, given by 
\begin{align}
    w^{j,h}_{k+1} \propto w^{j}_{k} e^{-\text{tr} \left((\mathbf{A}^{j,h})^{\text{T}}\mathbf{L}^{j}\right)}
\end{align}
subject to\footnote{To further reduce the complexity, we can prune those data associations with weights lower than a threshold, or only keep a certain number of data associations with top weights. If such methods are applied, weights should be renormalized.} $\sum_{j\in \mathbb{I}_{k}}\sum_{h \in \mathbb{H}^{j}_{k+1}} w^{j,h}_{k+1}=1$.

\subsubsection{Landmark Update} \label{Landmark_Update}

Under the $j,h$-th new data association $\boldsymbol{\sigma}^{j,h}$, we introduce the random variable  $\tilde{\boldsymbol{s}}^{j,h,i}_{k+1}=[\boldsymbol{s}^{\text{T}}_{k+1}, (\boldsymbol{x}^{j,h,i})^{\text{T}}]^{\text{T}}$ with mean $\tilde{\boldsymbol{m}}^{j,h,i}_{k+1|k}$,  comprising the predicted sensor state and the state of the $i$-th landmark under the $j,h$-th new data association, as well as the random variable 
$\check{\boldsymbol{s}}^{j,h}_{k+1}=[\boldsymbol{s}^{\text{T}}_{k+1},(\boldsymbol{x}^{j,h,1})^{\text{T}},\cdots,(\boldsymbol{x}^{j,h,{|\mathbb{I}_{k}^{j}|}})^{\text{T}}]^{\text{T}}$
 with mean $\check{\boldsymbol{m}}_{k+1|k}^{j,h}$.  
The mean vector, covariance matrix and measurement vector for the joint state of the sensor and all previously detected landmarks $i \in \mathbb{I}_{k+1}^{j,h},i\leq|\mathbb{I}_{k}^{j}|$ are constructed as 
\begin{align}
    &\check{\boldsymbol{m}}_{k+1|k}^{j,h}=[\boldsymbol{m}_{k+1|k}^{\text{T}},(\boldsymbol{u}^{j,h,1}_{k+1|k})^{\text{T}},\cdots,(\boldsymbol{u}^{j,h,|\mathbb{I}_{k}^{j}|}_{k+1|k})^{\text{T}}]^{\text{T}}, \label{joint_mean} \\&
    \check{\boldsymbol{P}}_{k+1|k}^{j,h}=\text{blkdiag}(\boldsymbol{P}_{k+1|k},\boldsymbol{C}^{j,h,1}_{k+1|k},\cdots,\boldsymbol{C}^{j,h,|\mathbb{I}_{k}^{j}|}_{k+1|k}), \label{joint_cov} \\&
    \check{\boldsymbol{h}}(\check{\boldsymbol{s}}^{j,h}_{k+1})=[\boldsymbol{h}(\tilde{\boldsymbol{s}}^{j,h,1}_{k+1})^{\text{T}},\cdots,\boldsymbol{h}(\tilde{\boldsymbol{s}}^{j,h,|\mathbb{I}_{k}^{j}|}_{k+1})^{\text{T}}]^{\text{T}}, \label{joint_h} \\&
    \hat{\boldsymbol{z}}_{k+1|k}^{j,h}=\check{\boldsymbol{h}}(\check{\boldsymbol{m}}^{j,h}_{k+1|k})
    % =\check{\boldsymbol{h}}_{k+1|k}^{j}(\check{\boldsymbol{m}}_{k+1|k}^{j,h})
    = \label{joint_inv}
    % \\& \qquad \quad
    [\boldsymbol{h}(\tilde{\boldsymbol{m}}^{j,h,1}_{k+1|k})^{\text{T}},\cdots,\boldsymbol{h}(\tilde{\boldsymbol{m}}^{j,h,|\mathbb{I}_{k}^{j}|}_{k+1|k})^{\text{T}}]^{\text{T}}, 
    % \nonumber 
    \\&
   % \check{\boldsymbol{z}}_{k+1}^{j,h}=[(\boldsymbol{z}^{j,h,\sigma^{j,h}(1)}_{k+1})^{\text{T}},\cdots,(\boldsymbol{z}^{j,h,\sigma^{j,h}(|\mathbb{I}_{k}^{j}|)}_{k+1})^{\text{T}}]^{\text{T}},\label{joint_meas}
    \check{\boldsymbol{z}}_{k+1}^{j,h}=[(\boldsymbol{z}^{\sigma^{j,h}(1)}_{k+1})^{\text{T}},\cdots,(\boldsymbol{z}^{\sigma^{j,h}(|\mathbb{I}_{k}^{j}|)}_{k+1})^{\text{T}}]^{\text{T}},\label{joint_meas}
\end{align}
where for $i: \sigma^{j,h}(i)=0$, we set $\boldsymbol{z}^{\sigma^{j,h}(i)}_{k+1}=\boldsymbol{0}$ and $\boldsymbol{h}(\tilde{\boldsymbol{m}}^{j,h,i}_{k+1|k})=\boldsymbol{0}$. % is the measurement associated to the landmark $\boldsymbol{x}^{j,i}$ under the $j,h$th new data association:
%\begin{align}
%    \boldsymbol{z}^{j,h,\sigma^{j,h}(i)}_{k+1} = \begin{cases}
 %   \boldsymbol{z}^{p}_{k+1} & \sigma^{j,h}(i)=p>0\\
  %  \boldsymbol{0} & \sigma^{j,h}(i)=0.
   % \end{cases}
%\end{align}
%which is $\boldsymbol{z}^{p}_{k+1}$ if $\sigma^{j,h}(i)=p>0$, and $\boldsymbol{0}_{|\boldsymbol{z}^{p}_{k+1}|\times1}$ if $\sigma^{j,h}(i)=0$. We also set $\boldsymbol{h}(\tilde{\boldsymbol{m}}^{j,h,i}_{k+1|k})=\boldsymbol{0}_{|\boldsymbol{z}^{1}_{k+1}|\times1}$ manually, if $\sigma^{j,h}(i)=0$. 

Then, the posterior mean and covariance under the $j,h$-th new data association can be determined via the first-order EK filter \cite[Ch. 5.2]{sarkka2013bayesian}
\begin{align}
    &\check{\boldsymbol{S}}_{k+1|k+1}^{j,h}=\check{\boldsymbol{H}}_{k+1|k}^{j}\check{\boldsymbol{P}}_{k+1|k}^{j,h}(\check{\boldsymbol{H}}_{k+1|k}^{j,h}
    )^{\text{T}} + \check{\boldsymbol{R}}_{k+1|k+1}^{j,h}, \label{joint_S}%\nonumber 
    \\&
    \boldsymbol{K}_{k+1|k+1}^{j,h}= \check{\boldsymbol{P}}_{k+1|k}^{j,h} (\check{\boldsymbol{H}}_{k+1|k}^{j,h})^{\text{T}} (\check{\boldsymbol{S}}_{k+1|k+1}^{j,h})^{-1}, %\nonumber 
    \\&
    \check{\boldsymbol{m}}_{k+1|k+1}^{j,h} = \check{\boldsymbol{m}}_{k+1|k}^{j,h}+\boldsymbol{K}_{k+1|k+1}^{j,h}(\check{\boldsymbol{z}}_{k+1|k+1}^{j,h}-\hat{\boldsymbol{z}}_{k+1|k}^{j,h}),% \nonumber 
    \\&
    \check{\boldsymbol{P}}_{k+1|k+1}^{j,h}= \check{\boldsymbol{P}}_{k+1|k}^{j,h} - \boldsymbol{K}_{k+1|k+1}^{j,h} \check{\boldsymbol{S}}_{k+1|k+1}^{j,h} (\boldsymbol{K}_{k+1|k+1}^{j,h})^{\text{T}},
    \\&
    %\check{\boldsymbol{R}}_{k+1|k+1}^{j,h}=\text{blkdiag}(\boldsymbol{R}^{j,h,\sigma^{j,h}(1)}_{k+1},\cdots,\boldsymbol{R}^{j,h,\sigma^{j,h}(\mathbb{I}_{k}^{j})}_{k+1}),\label{joint_R}
    \check{\boldsymbol{R}}_{k+1|k+1}^{j,h}=\text{blkdiag}(\boldsymbol{R}^{\sigma^{j,h}(1)}_{k+1},\cdots,\boldsymbol{R}^{\sigma^{j,h}(|\mathbb{I}_{k}^{j}|)}_{k+1}),\label{joint_R}
\end{align}
where we define $\boldsymbol{R}^{0}_{k+1} =  \boldsymbol{I}$. % if $\sigma^{j,h}(i)=p>0$, and $\boldsymbol{I}_{|\boldsymbol{z}^{1}_{k+1}|\times|\boldsymbol{z}^{1}_{k+1}|}$ if $\sigma^{j,h}(i)=0$, 
The matrix  $\check{\boldsymbol{H}}_{k+1|k}^{j,h}$ represents the Jacobian of $\check{\boldsymbol{h}}(\check{\boldsymbol{s}}^{j,h}_{k+1})$:
\begin{align}
    \check{\boldsymbol{H}}_{k+1|k}^{j,h} =\left. \frac{\partial \check{\boldsymbol{h}}(\check{\boldsymbol{s}}^{j,h}_{k+1})}{\partial \check{\boldsymbol{s}}_{k+1}^{j,h}}\right|_{\check{\boldsymbol{s}}_{k+1}^{j,h}=\check{\boldsymbol{m}}_{k+1|k}^{j,h}}.
\end{align}
%with elements 
%\begin{align}
 %   [\check{\boldsymbol{H}}_{k+1|k}^{j,h}]_{\alpha,\beta}=\left.\frac{\partial [\check{\boldsymbol{h}}_{k+1|k}^{j,h}(\check{\boldsymbol{s}}_{k+1|k}^{j,h})]_{\alpha}}{\partial [\check{\boldsymbol{s}}_{k+1|k}^{j,h}]_{\beta}}\right|_{\check{\boldsymbol{s}}_{k+1|k}^{j,h}=\check{\boldsymbol{m}}_{k+1|k}^{j,h}}.
%\end{align}
The mean and the covariance of the $j,h,i$-th landmark, for $i \in \mathbb{I}_{k+1}^{j,h},i\leq|\mathbb{I}_{k}^{j}|$, can be obtained from $\check{\boldsymbol{m}}_{k+1|k+1}^{j,h}$ and  $\check{\boldsymbol{P}}_{k+1|k+1}^{j,h}$ by extracting the corresponding parts of the posterior mean $\check{\boldsymbol{m}}_{k+1|k+1}^{j,h}$ and blocks along the diagonal of $\check{\boldsymbol{P}}_{k+1|k+1}^{j,h}$. %%%%The updated existence probabilities are given by \cite[Theorem 2]{williams2015marginal}
%marginalizing the other states out,  denoted as  
The updated existence probability is given by
\begin{align}
%    &\boldsymbol{u}^{j,h,i}_{k+1|k+1}= [\check{\boldsymbol{m}}_{k+1|k+1}^{j,h}]_{1+\nu + \mu(i-1) : \nu + \mu i}, \label{update_mean} \\&
 %   \boldsymbol{C}^{j,h,i}_{k+1|k+1}= [\check{\boldsymbol{P}}_{k+1|k+1}^{j,h}]_{1+\nu + \mu(i-1) : \nu + \mu i,1+\nu + \mu(i-1) : \nu + \mu i}, \label{update_covariance}\\&
    r^{j,h,i}_{k+1|k+1}=
    \begin{cases}
    1\quad& \sigma^{j,h}(i)>0  \\ \frac{(1-p_{\text{D},k+1}^{j,i})r^{j,i,0}_{k+1|k}}{1-r_{k+1|k}^{j,i,0}+(1-p_{\text{D},k+1}^{j,i})r_{k+1|k}^{j,i,0}} \quad & \sigma^{j,h}(i)=0,
    \end{cases}\label{update_r}
\end{align}
%where $\nu$ is the length of the $\boldsymbol{s}$, and $\mu$ is the length of $\boldsymbol{x}$. 

So-far, we have only considered the case of previously detected landmarks $i \in \mathbb{I}_{k+1}^{j,h},i\leq|\mathbb{I}_{k}^{j}|$. To account for the newly detected landmarks, we must also consider   $i \in \mathbb{I}_{k+1}^{j,h}, i>|\mathbb{I}_{k}^{j}|$, where $\sigma^{j,h}(\tilde{i})=p$ for some $p$, with $\sigma^{j,h}(\tilde{i})$ is the $i$-th non-empty component\footnote{For example, $\boldsymbol{\sigma}^{j,h}=[1,0,\emptyset,2]$ can form a MB with 3 Bernoullis, with the first two corresponding to the previously detected landmarks which is detected again with measurement 1 and misdetected, respectively. The third corresponds to the new Bernoulli detected with measurement 2. The new Bernoulli detected with measurement 1 do not exist. Therefore, $\sigma^{j,h}(\tilde{3})$ is the third non-empty (fourth) component in $\sigma^{j,h}$, which is 2.} in $\boldsymbol{\sigma}^{j,h}$. The corresponding posterior distributions do not affect the sensor state posterior and are given by 
\begin{align}
    &\boldsymbol{u}^{j,h,i}_{k+1|k+1}=\boldsymbol{u}^{p}_{\text{B},k+1|k+1}, \label{update_mean_newly_detected} \\&
    \boldsymbol{C}^{j,h,i}_{k+1|k+1}= \boldsymbol{C}^{p}_{\text{B},k+1|k+1}, \label{update_covariance_newly_detected} \\&
    r^{j,h,i}_{k+1|k+1}=\rho_{\text{B},k+1|k+1}^{p}/l_{\text{B},k+1}^{p},\label{update_r_newly_detected}
\end{align}
which were already computed in Appendix \ref{app:caseC} and Appendix \ref{app:birth}, when determining the local association weight in \eqref{weightnewdetected}.

\subsubsection{Sensor State Update} \label{sensor_update}
The vehicle mean and covariance, $\boldsymbol{m}_{k+1|k+1}$ and  $\boldsymbol{P}_{k+1|k+1}$, can be obtained from $\check{\boldsymbol{m}}_{k+1|k+1}^{j,h}$ and  $\check{\boldsymbol{P}}_{k+1|k+1}^{j,h}$ by marginalizing the landmark states out over all data associations:
\begin{align}\label{update_mean_covariance_vehicle}
    &\boldsymbol{m}_{k+1|k+1}= \sum_{j \in  \mathbb{I}_{k}} \sum_{h \in  \mathbb{H}^{j}_{k}} w_{k+1}^{j,h}[ \check{\boldsymbol{m}}_{k+1|k+1}^{j,h}]_{1:\nu} \\&
    \boldsymbol{P}_{k+1|k+1}= \sum_{j \in  \mathbb{I}_{k}}\sum_{h \in  \mathbb{H}^{j}_{k+1}} w_{k+1}^{j,h}([ \check{\boldsymbol{P}}_{k+1|k+1}^{j,h}]_{1:\nu,1:\nu}+ \\& \quad ([ \check{\boldsymbol{m}}_{k+1|k+1}^{j,h}]_{1:\nu}-\boldsymbol{m}_{k+1|k+1})([ \check{\boldsymbol{m}}_{k+1|k+1}^{j,h}]_{1:\nu}-\boldsymbol{m}_{k+1|k+1})^{\text{T}})\nonumber,
\end{align}
where $\nu$ is the length of the sensor state.

\subsubsection{Final Form after Update}
After vehicle and map update, 
the vehicle posterior distribution is ${\cal N}(\boldsymbol{s}_{k+1} ; \boldsymbol{m}_{k+1|k+1},\boldsymbol{P}_{k+1|k+1})$, while the map follows the PMBM format, with MBM components as $\{\{w_{k+1}^{j,h},\{r_{k+1|k+1}^{j,h,i},\boldsymbol{u}_{k+1|k+1}^{j,h,i},\boldsymbol{C}^{j,h,i}_{k+1|k+1} \}_{i\in \mathbb{I}_{k+1}^{j,h}}\}_{h\in \mathbb{H}^{j}_{k+1}}\}_{j\in \mathbb{I}_{k}}$ and PPP intensity as $\eta_{k+1|k+1}=(1-p_\text{D})\eta_{k+1|k}$, representing the previous undetected landmarks that remain undetected, with $p_\text{D}$ is a constant.

All data associations can be represented by only using one index. Hence, we reorder all data associations using index set $ \mathbb{I}_{k+1}=\{1,\cdots,\sum_{j\in \mathbb{I}_{k}}|\mathbb{H}^{j}_{k+1}|\}$. Then, MBM components can also be written as $\{w_{k+1}^{j},\{r_{k+1|k+1}^{j,i},\boldsymbol{u}_{k+1|k+1}^{j,i},\boldsymbol{C}^{j,i}_{k+1|k+1} \}_{i\in \mathbb{I}_{k+1}^{j}}\}_{j\in \mathbb{I}_{k+1}}$.

\section{Proposed EK-PMB SLAM Filter} \label{EKF-PMB}
The EK-PMBM SLAM filter generates $\gamma$ best global hypotheses for each prior global hypothesis. This means that the complexity of the EK-PMBM filter scales as $\mathcal{O}(|\bar{\mathbb{I}}|\gamma^{k-1})$ at time step $k$, where $|\bar{\mathbb{I}}|$ is the average number of Bernoullis over data associations. To avoid exponential complexity with time, we propose a variant based on the PMB filter, which only keeps one hypothesis at each time step. 

\subsection{Form of the EK-PMB Filter}
\sloppy In a PMB SLAM filter,
%, at time step $k$, the vehicle state 
%is given, approximated with a Gaussian distribution ${\cal N}(\boldsymbol{s}_{k|k} ; \boldsymbol{m}_{k|k},\boldsymbol{P}_{k|k})$, where $\boldsymbol{m}_{k|k}$ and $\boldsymbol{P}_{k|k}$ are the mean and the covariance matrix, respectively; 
the RFS landmark density comprises a PPP  $\eta_{k|k}$ and a MB  $\{r_{k|k}^{1,i},f_{k|k}^{1,i}(\boldsymbol{x})\}_{i\in \mathbb{I}^{1}_{k}}$, % for the map are also given, 
where each $f_{k|k}^{1,i}(\boldsymbol{x})$ is a  Gaussian distribution ${\cal N}(\boldsymbol{x}^{1,i}; \boldsymbol{u}_{k|k}^{1,i},\boldsymbol{C}^{1,i}_{k|k})$. % where $\boldsymbol{u}_{k|k}$ and $\boldsymbol{C}^{i}_{k|k}$ are the mean and the covariance matrix, respectively. 
Therefore, the MB components can also be written as  $\{r_{k|k}^{1,i},\boldsymbol{u}_{k|k}^{1,i},\boldsymbol{C}^{1,i}_{k|k} \}_{i\in \mathbb{I}_{k}^{1}}$. If we apply the update step from the EK-PMBM filter from Section \ref{EKF-PMBM}, the resulting map density at time step $k+1$ will be a PMBM with $\{w_{k+1}^{j},\{r_{k+1|k+1}^{j,i},\boldsymbol{u}_{k+1|k+1}^{j,i},\boldsymbol{C}^{j,i}_{k+1|k+1} \}_{i\in \mathbb{I}_{k+1}^{j}}\}_{j\in \mathbb{I}_{k+1}}$. Hence, to keep the PMB format, we need to approximate the MBM to an MB, which we do through a modified TOMB/P algorithm. %The standard TOMB approach from Section  \ref{sec:PMBSLAM} is not applicable, as will be shown below.

\subsection{Proposed PMB Approximation}
We firstly extend all MBs in the posterior PMBM density to the same space, with index set $\mathbb{T}_{k+1}$, where 
$|\mathbb{T}_{k+1}|=|\mathbb{I}^{1}_{k}|+|\mathcal{Z}_{k+1}|$
%. As the data associations under the prior MB becomes new MBs, all data associations have the same size. We introduce the index set for components in $\boldsymbol{\sigma}^{j}_{k+1}$ as $\mathbb{T}_{k+1}$, where $|\mathbb{T}_{k+1}|=|\mathbb{I}^{1}_{k}|+|\mathcal{Z}_{k+1}|$. 
Then, we can rewrite the MBM part of the posterior PMBM as %based on  $\boldsymbol{\sigma}^{j}_{k+1}$ as
\begin{align}
    &f_{k+1|k+1,\text{MBM}}(\mathcal{X}|\mathcal{Z}_{1:k+1}) = \label{MBM_rewrite}  \sum_{j\in\mathbb{I}_{k+1}}w^{j}_{k+1}\sum_{\uplus_{t \in \mathbb{T}_{k+1}}  \mathcal{X}^{t}=\mathcal{X}}%number\\& \qquad \qquad 
    \prod_{t\in\mathbb{T}_{k+1}}{f}^{t,\sigma^{j}(t)}_{k+1|k+1}(\mathcal{X}^{t}|\boldsymbol{\sigma}^{j}). %\notag 
\end{align}
where the Bernoulli density ${f}^{t,\sigma^{j}(t)}_{k+1|k+1}(\mathcal{X}^{t}|\boldsymbol{\sigma}^{j})$ captures the following cases (i) $t\leq |\mathbb{I}^{1}_{k}| $ and  $\sigma^{j}(t)=0$ correspond to  a previously detected landmark that is misdetected, (ii) $t\leq |\mathbb{I}^{1}_{k}| $ and  $\sigma^{j}(t)=p$ to a previously detected landmark that is detected with the $p$-th measurement, (iii) $t>|\mathbb{I}^{1}_{k}|$, $\sigma^{j}_{k+1}(t)=t-|\mathbb{I}^{1}_{k}|$ to a newly detected landmark with the $(t-|\mathbb{I}^{1}_{k}|)$-th measurement, and (iv) $t>|\mathbb{I}^{1}_{k}|$,  $\sigma^{j}(t)=\emptyset$ to a Bernoulli does not exist (with ${f}^{t,\emptyset}_{k+1|k+1}(\mathcal{X}^{t})$  having 0 existence probability).

The Bernoulli is conditioned on the full association vector $\boldsymbol{\sigma}^{j}$, due to the joint update in \eqref{joint_S}--\eqref{joint_R}. Before we can apply the TOMB/P algorithm, we must remove this conditioning, to obtain Bernoullis that only depend on the local association, i.e., of the form $\hat{f}^{t,\sigma^{j}(t)}_{k+1|k+1}(\mathcal{X}^{t})$. We do this by averaging and representing $\hat{f}^{t,\sigma^{j}(t)}_{k+1|k+1}(\mathcal{X}^{t})$ by the existence probability,  mean  and covariance, computed as (for $t\leq |\mathbb{I}^{1}_{k}|$ and $q\in \{0,1,\ldots, |\mathcal{Z}_{k+1}|\} \cup \emptyset$)
\begin{align}
    &{r}^{t,q}_{k+1|k+1} \propto \sum_{j \in  \mathbb{I}_{k+1}:  \sigma^{j}(t) =q} w_{k+1}^{j} r_{k+1|k+1}^{j,\sigma^{j}(t)}(\boldsymbol{\sigma}^j) \label{r_MBM}\\& \boldsymbol{u}^{t,q}_{k+1|k+1}\propto \sum_{j \in  \mathbb{I}_{k+1}:  \sigma^{j}(t) =q} w_{k+1}^{j} \boldsymbol{u}_{k+1|k+1}^{j,\sigma^{j}(t)}(\boldsymbol{\sigma}^j) \label{mean_MBM}\\&
    \boldsymbol{C}^{t,q}_{k+1|k+1}\propto \sum_{j \in  \mathbb{I}_{k+1}:  \sigma^{j}(t) =q} w_{k+1}^{j}( \boldsymbol{C}_{k+1|k+1}^{j,\sigma^{j}(t)}(\boldsymbol{\sigma}^j)+\label{covariance_MBM} \\ & (\boldsymbol{u}_{k+1|k+1}^{j, \sigma^{j}(t)}(\boldsymbol{\sigma}^j)-\boldsymbol{u}^{t,q}_{k+1|k+1})( \boldsymbol{u}_{k+1|k+1}^{j, \sigma^{j}(t)}(\boldsymbol{\sigma}^j)-\boldsymbol{u}^{t,q}_{k+1|k+1})^{\text{T}}), \notag 
\end{align}
where all three terms are normalized by the  marginal probability $\beta^{t,q}$ for $q\in \{0,1,\ldots, |\mathcal{Z}_{k+1}|\} \cup \emptyset$,  given by
\begin{align}
    &\beta^{t,q} = \sum_{j \in  \mathbb{I}_{k+1}:  \sigma^{j}(t) =q} w_{k+1}^{j}.\label{marginal_probability}
\end{align}

For $t>|\mathbb{I}^{1}_{k}|$, newly detected objects already satisfy $\hat{f}^{t,q}_{k+1|k+1}(\mathcal{X}^{t})={f}^{t,q}_{k+1|k+1}(\mathcal{X}^{t}|\boldsymbol{\sigma}^{j})$ by design, so averaging has no effect. Marginal association probabilities are given by 
\begin{align}
    \beta^{t,t-|\mathbb{I}^{1}_{k}|} & = \sum_{j \in  \mathbb{I}_{k+1}: \sigma^{j}(t) = t-|\mathbb{I}^{1}_{k}|} w_{k+1}^{j},\label{marginal_probability_newdet}\\
    \beta^{t,\emptyset} &  = 1-\beta^{t,t-|\mathbb{I}^{1}_{k}|}.
\end{align}
% The  marginal probability of the new Bernoulli existence $p_{k+1}(q^{t}_{k+1}=t-|\mathbb{I}^{1}_{k}), t>|\mathbb{I}^{1}_{k}|$ is given by
%\begin{align}
 %   &p_{k+1}(q^{t}_{k+1}=t-|\mathbb{I}^{1}_{k}|) = \sum_{j \in  \mathbb{I}_{k+1}: \exists \sigma^{j}_{k+1}(t) = t-|\mathbb{I}^{1}_{k}|} w_{k+1}^{j},\label{marginal_probability}
%\end{align}
%and the new Bernoulli non-existence %$p_{k+1}(q^{t}_{k+1}=\emptyset), %=1-p_{k+1}(q^{t}_{k+1}=t-|\mathbb{I}^{1}_{k}|)$.
%
Note that computing these marginal associations is straightforward, when $\gamma$ is not too large.
Finally, densities $\hat{f}^{t,q}_{k+1|k+1}(\mathcal{X}^{t})$ and the marginal probabilities $\beta^{t,q}$ are now be used as an input for the standard TOMB/P method to form the new MB \cite[Fig.10]{williams2015marginal}. 

\subsection{Overview of the EK-PMB(M) SLAM Filters}
The proposed EK-PMB(M) SLAM filter is summarized as Algorithm \ref{alg:EKF-PMB(M)}. The EK-PMB SLAM filter generates $\gamma$ best global hypotheses every time step. This means that the complexity of the EK-PMB filter scales as $\mathcal{O}(|\mathbb{I}^{1}_{k-1}|\gamma)$ at time $k$.

\begin{algorithm}[t]
\caption{Proposed EK-PMB(M) SLAM filter} \label{alg:EKF-PMB(M)}
\begin{algorithmic}[1]
%\Require \parbox[t]{\dimexpr\linewidth- \algorithmicindent * 1}{$\boldsymbol{m}_{k|k},\boldsymbol{P}_{k|k},\lambda_{k|k}(\boldsymbol{x}),\{w^{j}_{k},\{r_{k|k}^{j,i},\boldsymbol{u}_{k|k}^{j,i},\boldsymbol{C}^{j,i}_{k|k} \}_{i\in \mathbb{I}_{k}^{j}}\}_{j\in\mathbb{I}_{k}}$; for PMBM or \HW{something} for PMB\strut}
\Require \parbox[t]{\dimexpr\linewidth- \algorithmicindent * 1}{$\boldsymbol{m}_{k|k},\boldsymbol{P}_{k|k}$ and PMB(M) at time $k$%$\{w^{j}_{k},\{r_{k|k}^{j,i},\boldsymbol{u}_{k|k}^{j,i},\boldsymbol{C}^{j,i}_{k|k} \}_{i\in \mathbb{I}_{k}^{j}}\}_{j\in\mathbb{I}_{k}}$; for PMBM or \HW{something} for PMB
\strut}
\Ensure $\boldsymbol{m}_{k+1|k+1}$, $\boldsymbol{P}_{k+1|k+1}$, and updated PMB(M) at time $k+1$ %$\lambda_{k+1|k+1}(\boldsymbol{x})$, and  $\{w^{j}_{k+1},\{r_{k+1|k+1}^{j,i},\boldsymbol{u}_{k+1|k+1}^{j,i},\boldsymbol{C}^{j,i}_{k+1|k+1} \}_{i\in \mathbb{I}_{k+1}^{j}}\}_{j\in \mathbb{I}_{k+1}}$ for PMBM or \HW{something for PMB}.
\State Sensor state prediction %Compute $\boldsymbol{m}_{k+1|k}$, $\boldsymbol{P}_{k+1|k}$, $\{w^{j}_{k},\{r_{k+1|k}^{j,i},\boldsymbol{u}_{k+1|k}^{j,i},\boldsymbol{C}^{j,i}_{k+1|k} \}_{i\in \mathbb{I}_{k}^{j}}\}_{j\in \mathbb{I}_{k}}$ 
(Section \ref{EKF-pred}) 
\For{$j\in\mathbb{I}_{k}$} \algorithmiccomment{Each previous global hypothesis}
\State Map prediction %Compute $\boldsymbol{m}_{k+1|k}$, $\boldsymbol{P}_{k+1|k}$, $\{w^{j}_{k},\{r_{k+1|k}^{j,i},\boldsymbol{u}_{k+1|k}^{j,i},\boldsymbol{C}^{j,i}_{k+1|k} \}_{i\in \mathbb{I}_{k}^{j}}\}_{j\in \mathbb{I}_{k}}$ 
(Section \ref{EKF-pred}) 
\State Construct cost matrix $\mathbf{L}^{j}_{k+1}$ (Section \ref{Data_Association_Metric}) 
\State Compute best $\gamma$ data associations (Section \ref{Best_DA})
\State Compute updated PMBM (Section \ref{Landmark_Update}) %$\{w^{j}_{k+1},\{r_{k+1|k+1}^{j,i},\boldsymbol{u}_{k+1|k+1}^{j,i},\boldsymbol{C}^{j,i}_{k+1|k+1} \}_{i\in \mathbb{I}_{k+1}^{j}}\}_{j\in \mathbb{I}_{k+1}}$, $\lambda_{k+1|k+1}(\boldsymbol{x})$, $\boldsymbol{m}_{k+1|k+1}$, $\boldsymbol{P}_{k+1|k+1}$ can be acquired as shown in Section \ref{Landmark_Update} and Section \ref{sensor_update};
\EndFor
\State Compute updated sensor density (Section \ref{sensor_update}) 
\If {EK-PMB}  \algorithmiccomment{Convert to PMB}
%\Statex the additional steps need to be done to approximate the resulting MBM to MB:
\State Express MBM as \eqref{MBM_rewrite}
\State Merge Bernoullis %with same local association 
\eqref{r_MBM}--\eqref{covariance_MBM}
\State Compute marginal association probabilities \eqref{marginal_probability}--\eqref{marginal_probability_newdet}
\State Apply the TOMB/P method \cite[Fig.10]{williams2015marginal}
\EndIf
\end{algorithmic}
\end{algorithm}

\section{Extension to Multiple Models} \label{ekf-mm-pmbm}

In the data association, each measurement is associated to a landmark. However, the type of the landmark is still unknown. As mentioned in Section \ref{State model}, we consider three different types of landmarks, \ac{bs}, VA, and SP. Therefore, the SLAM filter should not only be able to figure out the source of each measurement, but also can distinguish the type of the associated landmark. 
\subsection{Problem Description}

The presented EK-PMBM and EK-PMB filters were designed only for continuous state spaces. Following \cite{li2020multiple}, we introduce the mixed state space comprising the continuous state $\boldsymbol{x}$ and the discrete state $\xi \in \mathcal{A}=\{\text{BS},\text{VA},\text{SP}\}$. 
%To do this, we implement the multiple model PMB(M) (MM-PMB(M))  into the proposed SLAM filter \cite{li2020multiple}. Therefore, we denote the type of the landmark as $\xi \in \mathcal{A}$, where $\mathcal{A}$ is $\{\text{BS},\text{VA},\text{SP}\}$. 
The corresponding Bernoulli densities are given by 
\begin{align}
    f^{j,i}_{\mathrm{B}}(\mathcal{X}^{i})=
\begin{cases}
1-r^{j,i} \quad& \mathcal{X}^{i}=\emptyset \\ r^{j,i}f^{j,i}(\boldsymbol{x},\xi)\psi^{j,i,\xi} \quad & \mathcal{X}^{i}=\{[\boldsymbol{x},\xi]\}, \\ 0 \quad & \mathrm{otherwise}
\end{cases}
\end{align}
where
%Therefore, we define the Bernoulli that has a discrete and continuous part, then the spatial density becomes $ f^{j,i}(\boldsymbol{x},\xi)=f^{j,i}(\boldsymbol{x}|\xi)\psi^{j,i,\xi}$, where 
$\psi^{j,i,\xi}\in [0,1]$ represents the  probability that the type of the $j,i$-th landmark is $\xi$, with $\sum_{\xi \in \mathcal{A}} \psi_{k|k}^{j,i,\xi} =1$. Furthermore, we set the detection probability model dependent, denoted as $p_{\text{D}}(\boldsymbol{x},\boldsymbol{s}_{k},\xi)$. 

We now describe how the EK-PMBM and EK-PMB filters should be modified to account for the multiple models. 

\subsection{Required Modifications}
\subsubsection{PMBM and PMB Form}
We can rewrite the PMBM components as $\{\eta_{k|k}^{\xi}\}_{\xi \in \mathcal{A}}$ and MBM parameters $\{w_{k}^{j,i},\{r_{k|k}^{j,i},\{\psi_{k|k}^{j,i,\xi},\boldsymbol{u}_{k|k}^{j,i,\xi},\boldsymbol{C}^{j,i,\xi}_{k|k}\}_{\xi \in \mathcal{A}}\}_{i\in \mathbb{I}_{k}^{j}}\}_{j\in \mathbb{I}_{k}}$. 

\subsubsection{Prediction Step}
The prediction step can be reformulated as $\eta_{k+1|k}^{\xi}=\eta_{k|k}^{\xi}$, $r_{k+1|k}^{j,i}=r_{k|k}^{j,i}$, $\psi_{k+1|k}^{j,i,\xi}=\psi_{k|k}^{j,i,\xi}$, $\boldsymbol{u}_{k+1|k}^{j,i,\xi}=\boldsymbol{u}_{k|k}^{j,i,\xi}$, $\boldsymbol{C}_{k+1|k}^{j,i,\xi}=\boldsymbol{C}_{k|k}^{j,i,\xi}$.

\subsubsection{Computation of Data Association Metric}
Compared to \ref{Data_Association_Metric}, there is only a different state definition. Hence, when calculating \eqref{weight_detected}, \eqref{weightmisdetected}, and \eqref{weightnewdetected} in the Appendices, we marginalize over $\psi$ in addition to $\boldsymbol{x}$.% It is identical to the previous expression, with just a different state definition. Computation of best $\gamma$ data associations is identical to the Section \ref{Best_DA}.
%In the joint update process, \eqref{detected4}, \eqref{weightmisdetected}, \eqref{rhonewdetected} can be replaced as
%\begin{align}
%    &\rho_{k+1|k+1}^{j,i,p}=\sum_{\xi \in \mathcal{A}} p_{\text{D},\xi}\psi_{k+1|k}^{j,i,\xi}{\cal N}(\boldsymbol{z}^{p} ; \boldsymbol{h}(\tilde{\boldsymbol{m}}^{j,i,\xi}_{k+1|k}),\boldsymbol{S}^{j,i,\xi}_{k+1|k}),\\
%    &l_{k+1|k+1}^{j,i,0}= l_{k+1|k}^{j,i}(1-r_{k+1|k}^{j,i}+(1-\sum_{\xi \in \mathcal{A}} p_{\mathrm{D},\xi}\psi_{k+1|k}^{j,i,\xi})r_{k+1|k}^{j,i}),\\
%    &\rho_{\text{B},k+1|k+1}^{p}=\tilde{\eta}\sum_{\xi \in \mathcal{A}}p_{\text{D},\xi} {\cal N}(\boldsymbol{z}^{p} ;\boldsymbol{h}(\tilde{\boldsymbol{m}}^{p,\xi}_{\text{B},k+1|k}),\boldsymbol{S}^{p,\xi}_{\text{B},k+1|k}),
%\end{align}
%respectively, where $\tilde{\boldsymbol{m}}^{j,i,\xi}_{k+1|k},\boldsymbol{S}^{j,i,\xi}_{k+1|k},\tilde{\boldsymbol{m}}^{p,\xi}_{\text{B},k+1|k}$ and  $\boldsymbol{S}^{p,\xi}_{\text{B},k+1|k}$ are defined in the same way as in the section \ref{joint_user_map_update}, only replacing the landmark state with $\boldsymbol{x}^{j,i,\xi}_{k+1|k}$ or $\boldsymbol{x}^{p,\xi}_{k+1|k}$, $\tilde{\eta}=\int \lambda_{k+1|k}(\boldsymbol{x},\xi) \text{d}\boldsymbol{x}\text{d}\xi$.

\subsubsection{Update Step}

To perform the update, we must consider all possible landmark types. Hence, in \eqref{joint_mean}, \eqref{joint_cov}, \eqref{joint_h}, \eqref{joint_inv}, $\boldsymbol{u}^{j,h,i}_{k+1|k}$, $\boldsymbol{C}^{j,h,i}_{k+1|k}$, $\boldsymbol{h}(\tilde{\boldsymbol{s}}^{j,h,i}_{k+1})$, $\boldsymbol{h}(\tilde{\boldsymbol{m}}^{j,h,i}_{k+1|k})$ should be replaced by
\begin{align}
    &\boldsymbol{u}^{j,h,i}_{k+1|k} \to [(\boldsymbol{u}^{j,h,i,1}_{k+1|k})^{\text{T}},\cdots,(\boldsymbol{u}^{j,h,i,|\mathcal{A}|}_{k+1|k})^{\text{T}}]^{\text{T}}, \\
    & \boldsymbol{C}^{j,h,i}_{k+1|k} \to \text{blkdiag}(\boldsymbol{C}^{j,h,i,1}_{k+1|k},\cdots,\boldsymbol{C}^{j,h,i,|\mathcal{A}|}_{k+1|k}),\\
    &\boldsymbol{h}(\tilde{\boldsymbol{s}}^{j,h,i}_{k+1}) \to [ \boldsymbol{h}(\tilde{\boldsymbol{s}}^{j,h,i,1}_{k+1})^{\text{T}},\cdots,\boldsymbol{h}(\tilde{\boldsymbol{s}}^{j,h,i,|\mathcal{A}|}_{k+1})^{\text{T}}]^{\text{T}},\\
    & \boldsymbol{h}(\tilde{\boldsymbol{m}}^{j,h,i}_{k+1|k}) \to [\boldsymbol{h}(\tilde{\boldsymbol{m}}^{j,h,i,1}_{k+1|k})^{\text{T}},\cdots,\boldsymbol{h}(\tilde{\boldsymbol{m}}^{j,h,i,|\mathcal{A}|}_{k+1|k})^{\text{T}}]^{\text{T}},
\end{align}
respectively. In \eqref{joint_meas}, $\boldsymbol{z}^{\sigma^{j,h}(i)}_{k+1}$ is to replaced by $\boldsymbol{1}_{|\mathcal{A}|\times 1}\otimes\boldsymbol{z}^{\sigma^{j,h}(i)}_{k+1}$, and  in \eqref{joint_R} $\boldsymbol{R}^{\sigma^{j,h}(t)}_{k+1}$  is replaced by $ \boldsymbol{1}_{|\mathcal{A}|\times |\mathcal{A}|} \otimes\boldsymbol{R}^{\sigma^{j,h}(t)}_{k+1}$, since replicating the measurements leads to perfect correlation in the covariance matrix. 
The updates can then be performed as before to recover the joint state posterior. Births should be generated for each landmark type (except $\xi = \text{BS}$).

To account for the type probabilities in the posterior, $\psi_{k+1|k+1}^{j,h,i,\xi}$, we compute, for $i\leq|\mathbb{I}^{j}_{k}|$
\begin{align}
    &\psi_{k+1|k+1}^{j,h,i,\xi}\propto   
    \begin{cases}
    (1-p_{\text{D},k+1}^{j,i,\xi}\psi_{k+1|k}^{j,i,\xi})& \sigma^{j,h}(i)=0  \\  p_{\text{D},k+1}^{j,i,\xi}\psi_{k+1|k}^{j,i,\xi}{\cal N}(\boldsymbol{z}_{k+1}^{p} ; \boldsymbol{h}(\tilde{\boldsymbol{m}}^{j,i,\xi}_{k+1|k}),\boldsymbol{S}^{j,i,\xi}_{k+1|k})  & \sigma^{j,h}(i)=p
    \end{cases}%\nonumber
\end{align}
where %$p_{\text{D}}(\tilde{\boldsymbol{s}}_{k+1|k}^{j,i,\xi})$ is assumed to be a constant over $\tilde{\boldsymbol{s}}_{k+1|k}^{j,i,\xi}$, with the value 
$p_{\text{D},k+1}^{j,i,\xi}= p_{\text{D}}(\tilde{\boldsymbol{s}}_{k+1}^{j,i,\xi}=\tilde{\boldsymbol{m}}_{k+1|k}^{j,i,\xi})$, and for $i>|\mathbb{I}^{j}_{k}|$,
$\sigma^{j,h}(\tilde{i})\neq \emptyset$
\begin{align}
    &\psi_{k+1|k+1}^{j,h,i,\xi}=\psi_{\text{B},k+1|k+1}^{\sigma^{j,h}(\tilde{i}),\xi}\propto
     {\eta}^{\xi}_{k+1|k} p_{\text{D},k+1}^{\sigma^{j,h}(\tilde{i}),\xi} {\cal N}(\boldsymbol{z}^{\sigma^{j,h}(\tilde{i})} ;\boldsymbol{h}(\tilde{\boldsymbol{m}}^{\sigma^{j,h}(\tilde{i}),\xi}_{\text{B},k+1|k}),\boldsymbol{S}^{\sigma^{j,h}(\tilde{i}),\xi}_{\text{B},k+1|k})  
    %\nonumber
\end{align}
where %$i\leq|\mathbb{I}^{j}_{k}|$ for the first two entries and $i>|\mathbb{I}^{j}_{k}|$ for the third entry,
the proportionality constant can be recovered from  $\sum_{\xi}\psi_{k+1|k+1}^{j,h,i,\xi}=1$. We recall that $\tilde{i}$ is the index of $i$-th non-empty component in $\sigma^{j,h}$, and  the global hypotheses $j,h$ will be finally re-indexed with $j$.

The PPP is updated by $\eta_{k+1|k+1}^{\xi}=(1-p_{\text{D}}^{\xi})\eta_{k+1|k}^{\xi}$, with $p_{\text{D}}^{\xi}$ is a known constant. %Moreover, $r^{j,i}_{k+1|k+1}$ in \eqref{update_r_newly_detected} needs to be replaced by 
%\begin{align}
%    &r^{j,i}_{k+1|k+1}=(1-\sum_{\xi \in \mathcal{A}} p_{\mathrm{D},\xi}w_{k+1|k+1}^{j,i,\xi}))r^{(j,i)^*,0}_{k+1|k}\nonumber \\& \quad/((1-r_{k+1|k}^{(j,i)^*,0}+(1-\sum_{\xi \in \mathcal{A}} p_{\mathrm{D},\xi}w_{k+1|k}^{j,i,\xi}))r_{k+1|k+1}^{(j,i)^*,0})).\label{update_mean_covariance_newly_detected_MM}
%\end{align}
%Other details are the same as in the section \ref{Landmark_Update}, except that the indexes of the corresponding landmark state and covariance in \eqref{update_mean} and \eqref{update_covariance} are different, and marginalization should be done over $\psi$ in \eqref{update_r} and \eqref{update_r_newly_detected}.

\subsubsection{PMB Approximation}
When approximating PMBM to PMB in Section \ref{EKF-PMB}, the landmark type should also be considered and %. Then, \eqref{r_MBM},
\eqref{mean_MBM}--\eqref{covariance_MBM} should be computed for each landmark type. Similarly, the marginal association probabilities must be computed for each landmark type by marginalizing the probability of the type for each association over data associations, for $t\leq|\mathbb{I}^{1}_{k}|$, to yield 
\begin{align}
\beta^{t,q,\xi}   \propto   
    \sum_{j \in  \mathbb{I}_{k+1}:  \sigma^{j}(t) =q} w_{k+1}^{j}\psi_{k+1|k+1}^{j,t,\xi},
\end{align}
and for $t>|\mathbb{I}^{1}_{k}|$,
\begin{align}
    \beta^{t,t-|\mathbb{I}^{1}_{k}|,\xi} \propto \sum_{j \in  \mathbb{I}_{k+1}:  \sigma^{j}(t) =t-|\mathbb{I}^{1}_{k}|} w_{k+1}^{j}\psi_{\text{B},k+1|k+1}^{t-|\mathbb{I}^{1}_{k}|,\xi}
\end{align}
with $\sum_{\xi}\beta^{t,q,\xi}=\beta^{t,q}$. We also have $\psi^{t,q,\xi}_{k+1|k+1}=\beta^{t,q,\xi}$.
%\begin{align}
    %&r^{t,q^{t}_{k+1}}_{k+1|k+1} \propto  \sum_{j \in  \mathbb{I}_{k+1}: \exists \sigma^{j}_{k+1}(t) = q^{t}_{k+1}} w_{k+1}^{j} r_{k+1|k+1}^{j,i=t} \\& \boldsymbol{u}^{t,q^{t}_{k+1},\xi}_{k+1|k+1}\propto \sum_{j \in  \mathbb{I}_{k+1}: \exists \sigma^{j}_{k+1}(t) = q^{t}_{k+1}} w_{k+1}^{j} \boldsymbol{u}_{k+1|k+1}^{j,i=t,\xi}\\& \boldsymbol{C}^{t,q^{t}_{k+1},\xi}_{k+1|k+1}\propto \sum_{j \in  \mathbb{I}_{k+1}: \exists \sigma^{j}_{k+1}(t) = q^{t}_{k+1}}  w_{k+1}^{j}( \boldsymbol{C}_{k+1|k+1}^{j,i=t,\xi}+ \\&  \qquad  (\boldsymbol{u}_{k+1|k+1}^{t,q^{t}_{k+1},\xi}-  \boldsymbol{u}^{j,i=t,\xi}_{k+1|k+1})( \boldsymbol{u}_{k+1|k+1}^{t,q^{t}_{k+1},\xi}-\boldsymbol{u}^{j,i=t,\xi}_{k+1|k+1})^{\text{T}}),\nonumber\\
%    \psi^{t,q^{t}_{k+1},\xi}_{k+1|k+1} = \psi_{k+1|k+1}^{j,i=t,\xi}, \quad \forall 
%j \in  \mathbb{I}_{k+1} ,\label{mean_covariance_MBM_MM}
%\end{align}
%where the first three terms are normalized by  $p_{k+1}(q^{t}_{k+1})$, and we further define the marginal probability of each type of source as
%\begin{align}
%   p_{k+1}(q^{t}_{k+1},\xi)  &= \sum_{j \in  \mathbb{I}_{k+1}: \exists \sigma^{j}_{k+1}(t) = q^{t}_{k+1}} w_{k+1}^{j}\psi_{k+1|k+1}^{j,i=t,\xi}\nonumber\\
%    &= \psi^{t,q^{t}_{k+1},\xi}_{k+1|k+1}p_{k+1}(q^{t}_{k+1}),\label{marginal_probability}
%\end{align}
%therefore, $\sum_{\xi \in \mathcal{A}} p_{k+1}(q^{t}_{k+1},\xi)=p_{k+1}(q^{t}_{k+1})$.
Then, we modify the TOMB/P algorithm \cite[Fig.10]{williams2015marginal} as shown in Algorithm \ref{alg:MMTOMB}.

\begin{algorithm}[t]
\caption{Modified MM-TOMB/P algorithm}% for approximating MM-MBM to MM-MB} 
\label{alg:MMTOMB}
\begin{algorithmic}[1]
%\KwData{this text}
\Require \parbox[t]{\dimexpr\linewidth- \algorithmicindent * 1}{Marginal probabilities $\beta^{t,q}$ and  $\beta^{t,q,\xi}$; MBM components;\strut}
\Ensure MB density
\begin{align*}
    \{\hat{r}^{1,i}_{k+1|k+1}, \{\hat{\psi}_{k+1|k+1}^{1,i,\xi},\hat{\boldsymbol{u}}_{k+1|k+1}^{1,i,\xi},\hat{\boldsymbol{C}}^{1,i,\xi}_{k+1|k+1}\}_{\xi \in \mathcal{A}}\}_{i\in\mathbb{T}_{k+1}}
\end{align*}%$\{r^{1,i}_{k+1|k+1}$, %$\{\psi_{k+1|k+1}^{1,i,\xi},\boldsymbol{u}_{k+1|k+1}^{1,i,\xi},\boldsymbol{C}^{1,i,\xi}_{k+1|k+1}\}_{\xi \in \mathcal{A}}\}_{i\in\mathbb{I}_{k+1}=\mathbb{T}_{k+1}}$.
%\State Form previously detected tracks
\For{$t \leq |\mathbb{I}^{1}_{k}|$} \algorithmiccomment{Previously detected tracks}
\State $\hat{r}^{1,t}_{k+1|k+1}=\sum_{q} \beta^{t,q} 
r^{t,q}_{k+1|k+1}$
\For{$\xi \in \mathcal{A}$}
\State $\hat{\psi}_{k+1|k+1}^{1,t,\xi}=\frac{\sum_{q} \beta^{t,q,\xi}r^{t,q}_{k+1|k+1}}{\hat{r}^{1,t}_{k+1|k+1}}$
\State $\hat{\boldsymbol{u}}_{k+1|k+1}^{1,t,\xi}=\frac{\sum_{q} \beta^{t,q,\xi}r^{t,q}_{k+1|k+1}\boldsymbol{u}_{k+1|k+1}^{t,q,\xi}}{\sum_{q} \beta^{t,q,\xi}r^{t,q}_{k+1|k+1}}$
\State $\hat{\boldsymbol{C}}_{k+1|k+1}^{1,i,\xi}= \frac{\sum_{q}\beta^{t,q,\xi}r^{t,q}_{k+1|k+1}(\boldsymbol{C}_{k+1|k+1}^{t,q,\xi}+(\hat{\boldsymbol{u}}_{k+1|k+1}^{1,t,\xi}-\boldsymbol{u}_{k+1|k+1}^{t,q,\xi})(\hat{\boldsymbol{u}}^{1,t,\xi}-\boldsymbol{u}_{k+1|k+1}^{t,q,\xi})^{\text{T}}) }{\sum_{q} \beta^{t,q,\xi}r^{t,q}_{k+1|k+1}}$
\EndFor
\EndFor
\For{$t > |\mathbb{I}^{1}_{k}|$} \algorithmiccomment{New tracks}
\State $\hat{r}^{1,t}_{k+1|k+1}= \beta^{t,t-|\mathbb{I}^{1}_{k}|} r^{t,t-|\mathbb{I}^{1}_{k}|}_{k+1|k+1}$
\For{$\xi \in \mathcal{A}$}
\State $\hat{\psi}_{k+1|k+1}^{1,t,\xi}=\frac{ \beta^{t,t-|\mathbb{I}^{1}_{k}|,\xi}r^{t,t-|\mathbb{I}^{1}_{k}|}_{k+1|k+1}}{ \hat{r}^{1,t}_{k+1|k+1}}$
\State $\hat{\boldsymbol{u}}_{k+1|k+1}^{1,t,\xi}=\boldsymbol{u}_{k+1|k+1}^{t,t-|\mathbb{I}^{1}_{k}|,\xi}$
\State $\hat{\boldsymbol{C}}_{k+1|k+1}^{1,t,\xi}= \boldsymbol{C}_{k+1|k+1}^{t,t-|\mathbb{I}^{1}_{k}|,\xi}$
\EndFor
\EndFor
\end{algorithmic}
\end{algorithm} 

\section{Results} \label{Results}
In this section, the proposed algorithms are evaluated on a vehicular scenario and compared to two benchmarks. We describe the simulation environment, the benchmarks and performance metrics, before discussing the SLAM results in terms of localization and mapping performance. %given scenario, and compared to the RBP-PMBM \cite{ge20205GSLAM} and EK-PHD \cite{EKPHD2021Ossi} SLAM filters. \Yu{To Henk: I feel that this sentence is a bit trivial, as we already have the Section VII.B. How do you think? }

\subsection{Simulation Environment} \label{sim_env}
\begin{figure}%[htbp]
\centerline{\includegraphics[width=0.5\linewidth]{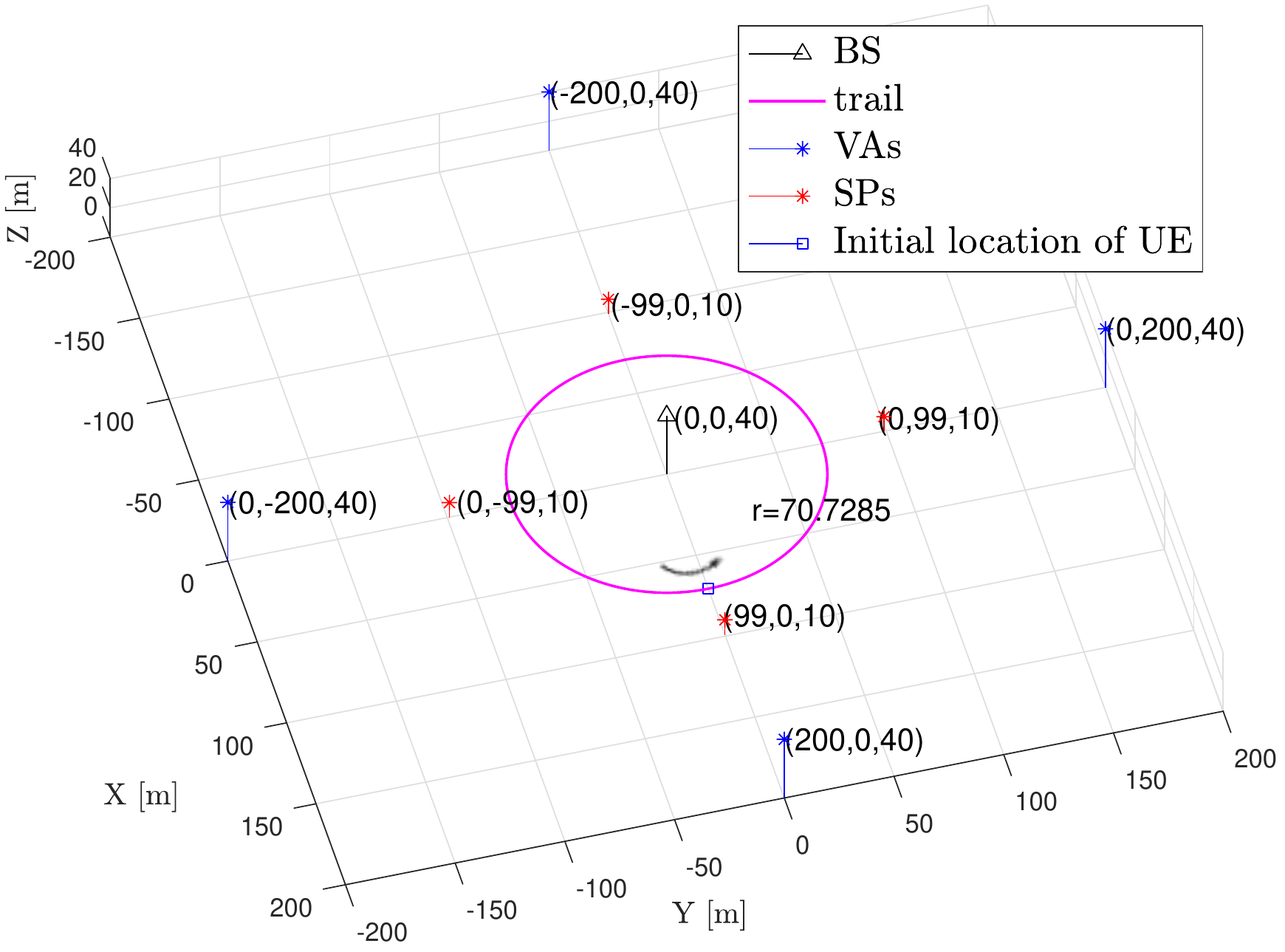}}
\caption{Scenario with the environment of a \ac{bs} and 4 VAs and 4 SPs. The \ac{ue} moves counterclockwise along the trail centered at the \ac{bs}.}
\label{fig:env}
  %\end{minipage}
\vspace{-4mm} \end{figure}

We consider a scenario as illustrated in Fig. \ref{fig:env}. There is a \ac{bs} located at $[0 \, \text{m},0 \, \text{m},40\, \text{m}]^{\mathrm{T}} $, 4 reflection surfaces with VAs located at $[ 200 \, \text{m},0 \, \text{m},40 \, \text{m}]^{\mathrm{T}}$, $[-200 \, \text{m}, 0 \, \text{m},40 \, \text{m}]^{\mathrm{T}}$, $[0 \, \text{m},200 \, \text{m},40 \, \text{m}]^{\mathrm{T}}$, $[0 \, \text{m}, -200 \, \text{m},40 \, \text{m}]^{\mathrm{T}}$, representing 4 reflection surfaces (wall in the physical environment), and 4 SPs, located at  $[99 \, \text{m},0 \, \text{m},10 \, \text{m}]^{\mathrm{T}}$, $[-99 \, \text{m},0 \, \text{m},10 \, \text{m}]^{\mathrm{T}}$, $[0 \, \text{m}, 99 \, \text{m},10 \, \text{m}]^{\mathrm{T}}$, $[0 \, \text{m}, -99 \, \text{m},10 \, \text{m}]^{\mathrm{T}}$, representing 
some small landmarks near the wall, for example some street lamps near the walls. In addition, there is a single vehicle doing a counterclockwise constant turn-rate movement around the \ac{bs}
 \begin{equation}
 \begin{split}
     &\boldsymbol{v}(\boldsymbol{s}_{k-1}) = \boldsymbol{s}_{k-1} + \left[
     \begin{array}{c}
	\frac{\zeta}{\varrho }(\sin{(\varpi_{k-1}+\varrho\Delta)}-\sin{\varpi_{k-1}})\\
	\frac{\zeta}{\varrho }(-\cos{(\varpi_{k-1}+\varrho\Delta)}+\cos{\varpi_{k-1}})\\
	0 \\
	\varrho\Delta\\
	0
	\end{array} \right],
\end{split}
\end{equation}
\sloppy where the state contains the position of the user $\boldsymbol{x}_{\mathrm{UE},k}=[x_{k},y_{k},z_{k}]^{\mathsf{T}}$, heading $\varpi_{k}$, and clock bias $B_{k}$,  $\zeta$ is the translation speed, set as $22.22 \, \text{m/s}$, $\varrho$ is the turn rate, set as $\pi/10 \, \text{rad/s}$, $\Delta$ is the sampling time interval, set as $0.5 \, \text{s}$. The covariance of the process noise $\boldsymbol{Q}$ is $\text{diag}[ 0.2 \, \text{m}^{2},0.2 \, \text{m}^{2},0 \, \text{m}^{2},0.001 \, \text{rad}^{2},0.2 \, \text{m}^{2}]$. The \ac{ue} is initialized at $[70.7285 \, \text{m},0 \, \text{m},0 \, \text{m},\pi/2 \, \text{rad}, 300 \, \text{m}]^{\mathrm{T}}$, and the initial covariance is $\text{diag}[ 0.3 \, \text{m}^{2},0.3 \, \text{m}^{2},0 \, \text{m}^{2},0.0052 \, \text{rad}^{2},0.3 \, \text{m}^{2}]$. 

To apply the proposed SLAM algorithm to the mmWave scenario, the Fisher information matrix of channel parameters~\cite{Zohair_5GFIM_TWC2018} is used to determine the measurement covariances at each time step. The Fisher information matrix is based on the following model. 
The \ac{bs} and the user are equipped with square planar arrays, and the numbers of \ac{bs} and user antennas are 64 and 16, respectively. The operating carrier frequency is 28~GHz. We consider the OFDM pilot signals with 16 symbols, 64 subcarriers, 200~MHz bandwidths. The transmitted power and noise spectral density are set to 5~dBm and -174~dBm/Hz, respectively. Path loss is generated according to \cite[eq.~(45)]{Zohair_5GFIM_TWC2018}, with reflection coefficient of surfaces set as 0.7, and the radar cross-section of SPs set as 50 $\text{m}^{2}$.

SPs are only visible when SPs are in the field-of-view of the \ac{ue}, which is set as $50 \, \text{m}$, while the \ac{bs} and VAs are always visible. The detection probabilities $p_{\mathrm{D},\xi}$ are set to 0.9 for \ac{bs}, VAs and visible SPs, and set to 0 for SPs which are out of \ac{fov}. The clutter intensity $c(\boldsymbol{z})$ is $1/(4\times200\pi^{4})$, with 200 representing the sensing range and 1 representing the average of the number of clutter measurements. %The birth intensities $\lambda_{\mathrm{B}}(\boldsymbol{x},\xi)$ are defined as $1.5 \times 10^{-5}$ for both VA and SP. 
Moreover, we utilize pruning, and merging for Bernoullis to decrease the computational complexity, with thresholds set as $10^{-4}$ and 50, respectively. 

\subsection{Studied Methods and Performance Metrics}
We evaluate the performance by comparing five algorithms: 
\begin{enumerate}
    \item the proposed EK-PMB SLAM filter with $\gamma=10$;
    \item the proposed EK-PMB SLAM filter with $\gamma=1$;
    \item the proposed EK-PMBM SLAM filter with $\gamma=10$;
    \item the RBP-PMBM SLAM filter \cite{ge20205GSLAM};
    \item the EK-PHD SLAM filter \cite{EKPHD2021Ossi}.
\end{enumerate}

%(1) the proposed EKF-PMB SLAM filter with $\gamma=10$; (2) the proposed EKF-PMB SLAM filter with $\gamma=1$, which is equivalent to doing the hard decision for Murty's algorithm; (3) \ac{rbpf} PMBM SLAM filter \cite{ge20205GSLAM}; (4) EKF-PHD SLAM filter \cite{EKPHD2021Ossi}, applied on the scenario proposed Section \ref{sim_env}. 

The mapping performance is measured by generalized optimal subpattern assignment (GOSPA) distance \cite{rahmathullah2017generalized}
\begin{align}
    &d_{\mathrm{GOSPA}}(\hat{\mathcal{X}},\mathcal{X})= \min_{\gamma\in\Gamma}(\sum_{(i,j)\in\gamma}d^{q_{p}}(\hat{\boldsymbol{x}}_{i},\boldsymbol{x}_{j})+\frac{q_{c}^{q_{p}}}{q_{a}}(N_{\mathrm{miss}}+N_{\mathrm{false}}))^{\frac{1}{q_{p}}},
\end{align}
where $\Gamma$ is the set of possible assignment set; $N_{\mathrm{miss}}$ is the number of miss detection; $N_{\mathrm{false}}$ is the number of false alarm. We set the cut-off distance $q_{c}=20$, the cardinality penalty factor $q_{a}=2$, the exponent factor $q_{p}=2$, and the measure accuracy of the state estimates is evaluated by the root mean squared error (RMSE) over time and the mean absolute error (MAE) changing with time. Overall, 10 Monte Carlo (MC) simulations and 1000 MC simulations are performed for the fourth and the rest algorithms, respectively, and the results are obtained by averaging over the different MC simulations. All the codes are written in MATLAB, and the simulations and experiments are run on a MacBook Pro (15-inch, 2019) with a 2.6 GHz 6-Core Intel Core i7 processor and 16 Gb memory.

\subsection{Results and Discussion}
Firstly, we study the mapping performance of the different SLAM methods. We observe that the first and third algorithms perform slightly better than the second algorithm, as the blue and black solid lines are lower than the red solid lines in Fig. \ref{Fig.mapping} and \ref{Fig.mapping_SP}. The reason is that the second algorithm takes the hard decision for the data association, which may pick up a wrong data association at some time steps, making the second algorithm not very stable and bringing additional error. The first and third algorithms perform similar, while the third algorithm has negligibly better performance. This is because even though the third algorithm keeps the PMBM format, there is usually a dominant MB, making the rest MBs unimportant. Moreover, from Fig. \ref{Fig.mapping} and Fig. \ref{Fig.mapping_SP}, we can find that the first four algorithms outperform  the EK-PHD SLAM filter, as the red dashed lines are higher than the others in both figures,  which is because the PHD filter cannot enumerate all possible data associations explicitly. Based on the PMBM filter, the RBP-PMBM filter is slightly better than the first three algorithms, as the nonlinearity is solved by using enough particles, and the density conditioned on each particle keeps the PMBM format and no approximation of MBM to MB is executed.
\begin{figure}
\center
\definecolor{mycolor1}{rgb}{0.00000,0.44700,0.74100}%
\definecolor{mycolor2}{rgb}{0.85000,0.32500,0.09800}%
\definecolor{mycolor3}{rgb}{0.00000,0.44700,0.74100}%
\definecolor{mycolor4}{rgb}{0.85000,0.32500,0.09800}%
\definecolor{mycolor5}{rgb}{0,0,0}%
%\definecolor{mycolor3}{rgb}{0.92900,0.69400,0.12500}%
%\definecolor{mycolor4}{rgb}{0.49400,0.18400,0.55600}%
%
\begin{tikzpicture}[scale=0.5\linewidth/14cm]

\begin{axis}[%
width=6.028in,
height=2.009in,
at={(1.011in,2.014in)},
scale only axis,
xmin=0,
xmax=40,
xlabel style={font=\color{white!15!black},font=\Large},
xlabel={time step},
ymin=0,
ymax=30,
ymode=log,
ylabel style={font=\color{white!15!black},font=\Large},
ylabel={GOSPA distance [m]},
axis background/.style={fill=white},
axis x line*=bottom,
axis y line*=left,
legend style={legend cell align=left, align=left, draw=white!15!black,font=\Large}
]
\addplot [color=mycolor1, line width=2.0pt]
  table[row sep=crcr]{%
1	28.2842712474619\\
2	14.2815076877367\\
3	2.3608900273532\\
4	1.86958621565284\\
5	1.84516225700986\\
6	1.82471616888289\\
7	1.80336156394962\\
8	1.7231975543124\\
9	1.61050436045507\\
10	1.47896855142484\\
11	1.37991171497453\\
12	1.31889459290686\\
13	1.27495049797053\\
14	1.19680768138537\\
15	1.14339323656076\\
16	1.1069822560735\\
17	1.07043255541158\\
18	1.06273699026761\\
19	1.07699476815325\\
20	1.01864494481351\\
21	1.01429663711902\\
22	0.987275820613174\\
23	0.977165012472249\\
24	0.970575564873686\\
25	0.954105283182213\\
26	0.937263354530606\\
27	0.914164646178292\\
28	0.889351685504464\\
29	0.882919032535644\\
30	0.864814996967124\\
31	0.845338189598242\\
32	0.840632351631002\\
33	0.839442372262749\\
34	0.821967612893701\\
35	0.822412161423321\\
36	0.818034945505461\\
37	0.804953291279147\\
38	0.793965542821192\\
39	0.771694398539424\\
40	0.768904263367888\\
};
\addlegendentry{EK-PMB with $\gamma=10$}

\addplot [color=mycolor2, line width=2.0pt]
  table[row sep=crcr]{%
1	28.2842712474619\\
2	14.61069038231981\\
3	2.19826893216357\\
4	2.38927711146112\\
5	2.13032718544272\\
6	1.94474480690431\\
7	1.97933545776544\\
8	1.77386528699347\\
9	1.6070924798458\\
10	1.50465853651786\\
11	1.41012643140318\\
12	1.38462024812301\\
13	1.32724705396888\\
14	1.27425011072173\\
15	1.22462386906783\\
16	1.19009341904979\\
17	1.14744627533138\\
18	1.15497554698392\\
19	1.14593421930795\\
20	1.11519096538842\\
21	1.07456747069257\\
22	1.07337321460548\\
23	1.07267204074422\\
24	1.05002654594153\\
25	1.04547118157319\\
26	1.03725850459664\\
27	1.02041671429084\\
28	1.02039692997218\\
29	1.01615805943879\\
30	1.00272215625274\\
31	0.992919026416869\\
32	0.983890333963561\\
33	0.966125092780923\\
34	0.960516638497756\\
35	0.947616401388431\\
36	0.936591470343951\\
37	0.913702551888904\\
38	0.905448113705986\\
39	0.89440618900431\\
40	0.891985854170959\\
};
\addlegendentry{EK-PMB with $\gamma=1$}

\addplot [color=mycolor5, line width=2.0pt]
  table[row sep=crcr]{%
1	28.2842712474619\\
2	14.4430664421268\\
3	2.01789900863084\\
4	1.75984004533113\\
5	1.78799162426577\\
6	1.77465707360237\\
7	1.79437876659862\\
8	1.59098915687467\\
9	1.44735577776576\\
10	1.32085506147508\\
11	1.2258595265087\\
12	1.17019999238103\\
13	1.09783332677967\\
14	1.074394284911376\\
15	0.982468827590847\\
16	0.974887982280766\\
17	0.959773489500317\\
18	0.912930092261505\\
19	0.896497976125273\\
20	0.915120828330582\\
21	0.905782483298157\\
22	0.884506918375166\\
23	0.852707193708485\\
24	0.848473909549964\\
25	0.830425134272638\\
26	0.830568921789277\\
27	0.811621905407414\\
28	0.800906983860668\\
29	0.788471541434889\\
30	0.766743547901116\\
31	0.761730214297476\\
32	0.766059131894456\\
33	0.778512660636621\\
34	0.75745049384834\\
35	0.741829362714729\\
36	0.740837984650758\\
37	0.732610289814097\\
38	0.731837948118093\\
39	0.725911880211699\\
40	0.730716498194207\\
};
\addlegendentry{EK-PMBM with $\gamma=10$}

\addplot [color=mycolor3, dashed, line width=2.0pt]
  table[row sep=crcr]{%
1	28.2842712474619\\
2	14.1803857775323\\
3	1.50812355513916\\
4	1.72277884814771\\
5	1.70831830238888\\
6	1.42406945946222\\
7	1.28794218772365\\
8	1.51416385612944\\
9	1.38268180621675\\
10	1.28371241249673\\
11	0.985651208176957\\
12	0.792895287022791\\
13	0.805932833188666\\
14	0.762946632949497\\
15	0.834267347792315\\
16	0.870129315123838\\
17	0.832483924910944\\
18	0.485210078939622\\
19	0.699260833372067\\
20	0.815166029897717\\
21	0.527902609866416\\
22	0.544960541854684\\
23	0.687760240404335\\
24	0.6526603300102\\
25	0.647619432477474\\
26	0.644751560978913\\
27	0.675031951875613\\
28	0.451323801265745\\
29	0.440465130885032\\
30	0.503362306760034\\
31	0.524372187058061\\
32	0.496057011411148\\
33	0.45095054468586\\
34	0.483495323333383\\
35	0.452424496806593\\
36	0.453239419863599\\
37	0.516779474802085\\
38	0.521324095236159\\
39	0.455142149449697\\
40	0.441573030627103\\
};
\addlegendentry{RBP-PMBM}

\addplot [color=mycolor4, dashed, line width=2.0pt]
  table[row sep=crcr]{%
1	28.2842712474619\\
2	21.9560780283641\\
3	7.54796845621767\\
4	7.76364840021301\\
5	5.45852711804379\\
6	4.40228915989216\\
7	4.69444305080753\\
8	3.81216841877973\\
9	3.12672216204169\\
10	3.140581393565\\
11	2.52032254937619\\
12	2.70488285522605\\
13	2.92636887563318\\
14	2.24577160977453\\
15	2.38945188087766\\
16	2.35094708498232\\
17	2.29582989022679\\
18	2.3143262790524\\
19	2.38298352397398\\
20	2.32916444927525\\
21	2.3270164183666\\
22	2.24105029379197\\
23	2.03313046552227\\
24	2.16430740470556\\
25	2.23666585164272\\
26	2.34738680895363\\
27	2.14219132977888\\
28	2.27943552257275\\
29	2.14384926234255\\
30	2.0016214396336\\
31	1.97579907931694\\
32	2.02628573567747\\
33	2.09604471631159\\
34	2.16139997443453\\
35	2.12140344376624\\
36	2.00159357359082\\
37	1.97686497667226\\
38	2.11211045066419\\
39	2.22050049651342\\
40	2.28902257403792\\
};
\addlegendentry{EK-PHD}

\end{axis}
\end{tikzpicture}%
\caption{Comparison of mapping performances for VAs among 5 algorithms.}
\label{Fig.mapping}
\vspace{-4mm} \end{figure}
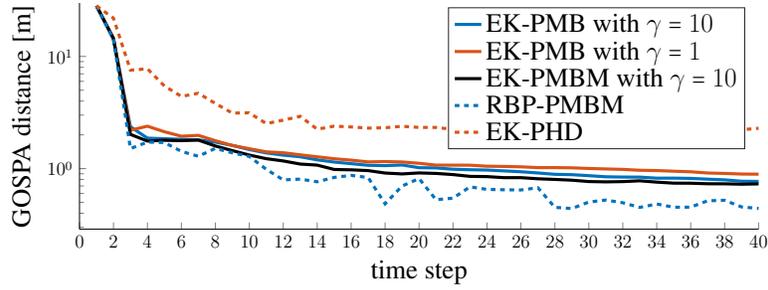
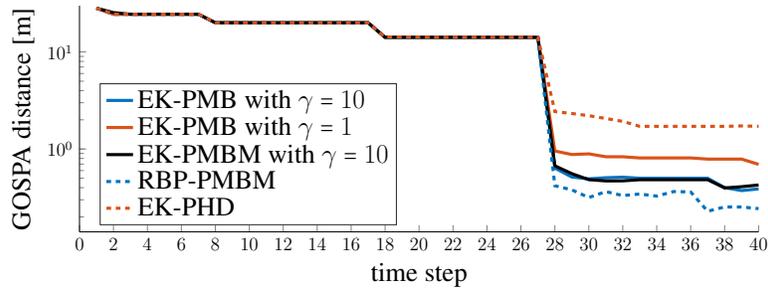
\begin{figure}
\center
\definecolor{mycolor1}{rgb}{0.00000,0.44700,0.74100}%
\definecolor{mycolor2}{rgb}{0.85000,0.32500,0.09800}%
\definecolor{mycolor3}{rgb}{0.00000,0.44700,0.74100}%
\definecolor{mycolor4}{rgb}{0.85000,0.32500,0.09800}%
\definecolor{mycolor5}{rgb}{0,0,0}%
%\definecolor{mycolor3}{rgb}{0.92900,0.69400,0.12500}%
%\definecolor{mycolor4}{rgb}{0.49400,0.18400,0.55600}%
%
\begin{tikzpicture}[scale=0.5\linewidth/14cm]

\begin{axis}[%
width=6.028in,
height=2.009in,
at={(1.011in,2.014in)},
scale only axis,
xmin=0,
xmax=40,
xlabel style={font=\color{white!15!black},font=\Large},
xlabel={time step},
ymin=0,
ymax=30,
ymode=log,
ylabel style={font=\color{white!15!black},font=\Large},
ylabel={GOSPA distance [m]},
axis background/.style={fill=white},
axis x line*=bottom,
axis y line*=left,
legend pos=south west,
legend style={legend cell align=left, align=left, draw=white!15!black,font=\Large}
]
\addplot [color=mycolor1, line width=2.0pt]
  table[row sep=crcr]{%
1	28.2842712474619\\
2	24.4976104126054\\
3	24.4971726800625\\
4	24.4971726800625\\
5	24.4971726800625\\
6	24.4971726800625\\
7	24.4971726800625\\
8	20.0089606935208\\
9	20.006432279241\\
10	20.0043857271464\\
11	20.0046877977079\\
12	20.0051328095331\\
13	20.0048052713288\\
14	20.0048052713288\\
15	20.0048052713288\\
16	20.0048052713288\\
17	20.0048052713288\\
18	14.1602334840715\\
19	14.1542186548884\\
20	14.1514978309904\\
21	14.1509867039574\\
22	14.1502552456764\\
23	14.1500692849031\\
24	14.1500692849031\\
25	14.1500692849031\\
26	14.1500692849031\\
27	14.1500692849031\\
28	0.637075649071575\\
29	0.514538379627771\\
30	0.492985276726793\\
31	0.505962405801442\\
32	0.512787304142804\\
33	0.500015157159195\\
34	0.500015157159195\\
35	0.500015157159195\\
36	0.500015157159195\\
37	0.499934267880708\\
38	0.400520670101598\\
39	0.373244048861521\\
40	0.388478532031204\\
};
\addlegendentry{EK-PMB with $\gamma=10$}

\addplot [color=mycolor2, line width=2.0pt]
  table[row sep=crcr]{%
  1	28.2842712474619\\
2	24.5064571081199\\
3	24.5013158304374\\
4	24.5013158304374\\
5	24.5013158304374\\
6	24.5013158304374\\
7	24.5013158304374\\
8	20.0150042013919\\
9	20.0131413894822\\
10	20.0119284496721\\
11	20.0109209274677\\
12	20.0125153216728\\
13	20.0126412925715\\
14	20.0126412925715\\
15	20.0126412925715\\
16	20.0126412925715\\
17	20.0126412925715\\
18	14.1735315745703\\
19	14.1704259925714\\
20	14.1647667323002\\
21	14.1606511961838\\
22	14.1606739238941\\
23	14.1609768707514\\
24	14.1609768707514\\
25	14.1609768707514\\
26	14.1609768707514\\
27	14.1609768707514\\
28	0.954050507105891\\
29	0.877818859544393\\
30	0.891589809903384\\
31	0.832712895320372\\
32	0.832155531989099\\
33	0.809488908553463\\
34	0.809488908553463\\
35	0.809488908553463\\
36	0.809488908553463\\
37	0.787730050703057\\
38	0.787730050703057\\
39	0.787730050703057\\
40	0.695162348415402\\
};
\addlegendentry{EK-PMB with $\gamma=1$}
\addplot [color=mycolor5, line width=2.0pt]
  table[row sep=crcr]{%
1	28.2842712474619\\
2	25.4450920267171\\
3	24.4976084237915\\
4	24.4976084237915\\
5	24.4976084237915\\
6	24.4976084237915\\
7	24.4976084237915\\
8	20.0116132309335\\
9	20.0075508016813\\
10	20.0055574989945\\
11	20.0056059010642\\
12	20.005459418142\\
13	20.0054723492403\\
14	20.0054723492403\\
15	20.0054723492403\\
16	20.0054723492403\\
17	20.0054723492403\\
18	14.1626121347146\\
19	14.153971121811\\
20	14.1532060110624\\
21	14.151508648118\\
22	14.1506709323724\\
23	14.150515747314\\
24	14.150515747314\\
25	14.150515747314\\
26	14.150515747314\\
27	14.150515747314\\
28	0.67158057945417\\
29	0.557490113070335\\
30	0.483406285972127\\
31	0.469356042946552\\
32	0.47006336947393\\
33	0.48282434207054\\
34	0.48282434207054\\
35	0.482824342070542\\
36	0.482824342070542\\
37	0.482824342070542\\
38	0.397302411898653\\
39	0.409103633320858\\
40	0.425969771023717\\
};
\addlegendentry{EK-PMBM with $\gamma=10$}

\addplot [color=mycolor3, dashed, line width=2.0pt]
  table[row sep=crcr]{%
1	28.2842712474619\\
2	24.4978826644068\\
3	24.4963723041687\\
4	24.4958725680683\\
5	24.4964735361943\\
6	24.4965770333216\\
7	24.4959647935387\\
8	20.0031058172215\\
9	20.0042873847625\\
10	20.0033419546934\\
11	20.0026327937978\\
12	20.0015246159575\\
13	20.0010730067171\\
14	20.0019723655925\\
15	20.0022157382861\\
16	20.0022080655755\\
17	20.0023120634136\\
18	14.1533124178684\\
19	14.1468694442103\\
20	14.1471326240372\\
21	14.1454474540427\\
22	14.1455653547584\\
23	14.1457198926402\\
24	14.1461038172215\\
25	14.1461496954345\\
26	14.1456867871757\\
27	14.145461478682\\
28	0.418497330109521\\
29	0.379603098644561\\
30	0.316375950268309\\
31	0.364009038554483\\
32	0.331511402470091\\
33	0.343940132150374\\
34	0.3274902107448\\
35	0.364743212592972\\
36	0.361051426457231\\
37	0.228475872945747\\
38	0.253156275080025\\
39	0.252748115195989\\
40	0.243095977559814\\
};
\addlegendentry{RBP-PMBM}

\addplot [color=mycolor4, dashed, line width=2.0pt]
  table[row sep=crcr]{%
1	28.2842712474619\\
2	24.5155827096825\\
3	24.5155827096825\\
4	24.5155827096825\\
5	24.5155827096825\\
6	24.5155827096825\\
7	24.5155827096825\\
8	20.015168159774\\
9	20.0098063880665\\
10	20.0086722683666\\
11	20.0028991727098\\
12	20.0042408656914\\
13	20.0042408656914\\
14	20.0042408656914\\
15	20.0042408656914\\
16	20.0042408656914\\
17	20.0042408656914\\
18	14.2721001040997\\
19	14.2568668319866\\
20	14.2196659194445\\
21	14.1660749947593\\
22	14.180384995944\\
23	14.180384995944\\
24	14.180384995944\\
25	14.180384995944\\
26	14.180384995944\\
27	14.180384995944\\
28	2.4299391434233\\
29  2.32327133095552\\
30	2.20451539580507\\
31	2.06375356005761\\
32	1.92752030833835\\
33	1.71314239090241\\
34	1.71314239090241\\
35	1.71314239090241\\
36	1.71314239090241\\
37	1.71314239090241\\
38	1.71314239090241\\
39	1.72934584634096\\
40	1.71436132982418\\
};
\addlegendentry{EK-PHD}

\end{axis}
\end{tikzpicture}%
\caption{Comparison of mapping performances for SPs among 5 algorithms.}
\label{Fig.mapping_SP}
\vspace{-4mm} \end{figure}

Next, performance  of  the  proposed SLAM filter in vehicle state estimation is studied. Fig. \ref{Fig.positioning} shows the RMSEs of the estimated vehicle position, bias, and heading of the five SLAM filters, and Fig. \ref{Fig.positioning_line} shows the MAE of the estimated position changing with time. Overall, all four PMB(M)-filter-based SLAM filters perform better than the EK-PHD SLAM filter, which is due to the PHD filter do not have explicit enumeration of the different data associations, and approximate the posterior density to a PPP. Within the four PMB(M)-filter-based SLAM filters, the RBP-SLAM filter performs the best, which is due to enough particles are used to solve the nonlinearity, all possible data associations are tracked to keep the PMBM format of the density conditioned on each particle. Because of not doing hard decision in data association and considering more than one MBs, the EK-PMB SLAM filter with $\gamma=10$ and EK-PMBM SLAM filter perform slightly better than the EK-PMB SLAM filter with $\gamma=1$. However, these two filters perform similar, as the sensor state prediction and update processes are the same in both algorithms, and there is usually one global hypothesis having the dominant weight every time step.
\begin{figure}
\center
\definecolor{mycolor1}{rgb}{0.00000,0.44700,0.74100}%
\definecolor{mycolor2}{rgb}{0.85000,0.32500,0.09800}%
\definecolor{mycolor3}{rgb}{0.92900,0.69400,0.12500}%
\definecolor{mycolor4}{rgb}{0.49400,0.18400,0.55600}
\definecolor{mycolor5}{rgb}{0,0,0}%
\begin{tikzpicture}[scale=0.6\linewidth/14cm]

\begin{axis}[%
width=3.842in,
height=1.281in,
at={(3.465in,2.378in)},
scale only axis,
bar shift auto,
xmin=0.5,
xmax=3.5,
xtick={1,2,3},
xticklabels={{position},{heading},{clock bias}},
ymin=0,
ymax=0.3,
ylabel style={font=\color{white!15!black}},
ylabel={RMSE of state [m], [deg], [m]},
axis background/.style={fill=white},
legend style={at={(-0.15,1)}, anchor=north east, legend cell align=left, align=left, draw=white!15!black}
]
\addplot[ybar, bar width=0.145, fill=mycolor1, draw=black, area legend] table[row sep=crcr] {%
1	0.1663\\
2	0.1362\\
3	0.0898\\
};
\addplot[forget plot, color=white!15!black] table[row sep=crcr] {%
0.5	0\\
3.5	0\\
};
\addlegendentry{EK-PMB with $\gamma=10$}

\addplot[ybar, bar width=0.145, fill=mycolor2, draw=black, area legend] table[row sep=crcr] {%
1	0.1979\\
2	0.1693\\
3	0.1016\\
};
\addplot[forget plot, color=white!15!black] table[row sep=crcr] {%
0.5	0\\
3.5	0\\
};
\addlegendentry{EK-PMB with $\gamma=1$}

\addplot[ybar, bar width=0.145, fill=mycolor5, draw=black, area legend] table[row sep=crcr] {%
1	0.1604\\
2	0.1339\\
3	0.0857\\
};
\addplot[forget plot, color=white!15!black] table[row sep=crcr] {%
0.5	0\\
3.5	0\\
};
\addlegendentry{EK-PMBM with $\gamma=10$}

\addplot[ybar, bar width=0.145, fill=mycolor3, draw=black, area legend] table[row sep=crcr] {%
1	0.1477\\
2	0.1048\\
3	0.0753\\
};
\addplot[forget plot, color=white!15!black] table[row sep=crcr] {%
0.5	0\\
3.5	0\\
};
\addlegendentry{RBP-PMBM}

\addplot[ybar, bar width=0.145, fill=mycolor4, draw=black, area legend] table[row sep=crcr] {%
1	0.2542\\
2	0.2078\\
3	0.1423\\
};
\addplot[forget plot, color=white!15!black] table[row sep=crcr] {%
0.5	0\\
3.5	0\\
};
\addlegendentry{EK-PHD}

\end{axis}
\end{tikzpicture}%
\caption{Comparison of vehicle state estimation among 5 algorithms.}
\label{Fig.positioning}
\vspace{-4mm} \end{figure}
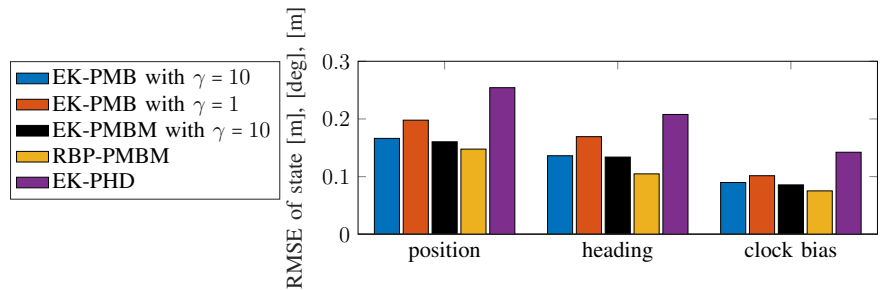

\begin{figure}
\center
\definecolor{mycolor1}{rgb}{0.00000,0.44700,0.74100}%
\definecolor{mycolor2}{rgb}{0.85000,0.32500,0.09800}%
\definecolor{mycolor3}{rgb}{0.92900,0.69400,0.12500}%
\definecolor{mycolor4}{rgb}{0.49400,0.18400,0.55600}%
\definecolor{mycolor5}{rgb}{0,0,0}
\begin{tikzpicture}[scale=0.5\linewidth/14cm]
\begin{axis}[%
width=6.028in,
height=2.009in,
at={(1.011in,2.014in)},
scale only axis,
xmin=0,
xmax=40,
xlabel style={font=\color{white!15!black},font=\Large},
xlabel={time step},
ymin=0,
ymax=0.5,
ymode=log,
ylabel style={font=\color{white!15!black},font=\Large},
ylabel={MAE of position [m]},
axis background/.style={fill=white},
axis x line*=bottom,
axis y line*=left,
legend pos=south west,
legend style={legend cell align=left, align=left, draw=white!15!black,font=\Large}
]
\addplot [color=mycolor1, line width=2.0pt]
  table[row sep=crcr]{%
1	0.244440000515429\\
2	0.205290109616914\\
3	0.221327431238034\\
4	0.20154519478018\\
5	0.202943276492121\\
6	0.179124525804273\\
7	0.186025990326865\\
8	0.1756059435913\\
9	0.19464951154528\\
10	0.189741132585282\\
11	0.19658152553044\\
12	0.197150248032116\\
13	0.185864478447547\\
14	0.188395916284782\\
15	0.200622668741351\\
16	0.19504336252507\\
17	0.196381738635173\\
18	0.179452460992472\\
19	0.19176767720128\\
20	0.198168058773379\\
21	0.162096670965647\\
22	0.174451459215614\\
23	0.18573836500874\\
24	0.167517667698087\\
25	0.157104401030743\\
26	0.157742915583309\\
27	0.168879731931658\\
28	0.134116529748964\\
29	0.152888885028984\\
30	0.121416765229553\\
31	0.123966301628717\\
32	0.133049006821639\\
33	0.134105464433002\\
34	0.135649616295674\\
35	0.13015730116451\\
36	0.115443192691654\\
37	0.129475338390738\\
38	0.103662606780303\\
39	0.118698337722304\\
40	0.114061248516018\\
};
\addlegendentry{EK-PMB with $\gamma=10$}

\addplot [color=mycolor2, line width=2.0pt]
  table[row sep=crcr]{%
1	0.265317869343472\\
2	0.230984376246761\\
3	0.22435271738992\\
4	0.224085023516584\\
5	0.234038625941373\\
6	0.223092554229443\\
7	0.232292948536294\\
8	0.236714782189392\\
9	0.242408079500099\\
10	0.250125736338311\\
11	0.235602986511245\\
12	0.230882521233124\\
13	0.231187327301165\\
14	0.234776745355362\\
15	0.22369677633864\\
16	0.224762548722461\\
17	0.223297755909402\\
18	0.199290920701279\\
19	0.199315186447989\\
20	0.211189005346257\\
21	0.17655234956308\\
22	0.194927789567335\\
23	0.180840766546827\\
24	0.184922075140926\\
25	0.204704279483\\
26	0.189259891156599\\
27	0.183633852374608\\
28	0.173067604130544\\
29	0.183859808574596\\
30	0.17642658882136\\
31	0.165277115022679\\
32	0.166672371123579\\
33	0.152567810431965\\
34	0.158587063664491\\
35	0.148807240973771\\
36	0.155970933466099\\
37	0.147246195342202\\
38	0.143565202466063\\
39	0.149149077301643\\
40	0.146044865766664\\
};
\addlegendentry{EK-PMB with $\gamma=1$}
\addplot [color=mycolor5, line width=2.0pt]
  table[row sep=crcr]{%
1	0.24605584179777\\
2	0.191761104686047\\
3	0.193217839853405\\
4	0.194972113416861\\
5	0.181436573165629\\
6	0.175147705869261\\
7	0.187713102612364\\
8	0.184073911782016\\
9	0.176021547540539\\
10	0.178433147829893\\
11	0.186755250664017\\
12	0.169098406892002\\
13	0.168662928309646\\
14	0.163194373895529\\
15	0.169637249546059\\
16	0.161854722109028\\
17	0.170459974512376\\
18	0.168126061655579\\
19	0.167546359465238\\
20	0.154388035389341\\
21	0.167374421381603\\
22	0.163537875781349\\
23	0.145595481382161\\
24	0.177057198993474\\
25	0.171247316862037\\
26	0.165907069742891\\
27	0.158795802366045\\
28	0.13974798660183\\
29	0.155828005997876\\
30	0.141490921119289\\
31	0.140516028676828\\
32	0.122100387651142\\
33	0.137107964680661\\
34	0.130957847896183\\
35	0.1302837188732\\
36	0.123923596346672\\
37	0.115529269277191\\
38	0.122036736495554\\
39	0.123884167685235\\
40	0.115714518558786\\
};
\addlegendentry{EK-PMBM with $\gamma=10$}
\addplot [color=mycolor1, dashed, line width=2.0pt]
  table[row sep=crcr]{%
1	0.261632147788967\\
2	0.182320520119313\\
3	0.0654916428932192\\
4	0.0626561610113271\\
5	0.163902515881458\\
6	0.155808702454556\\
7	0.0692633865181821\\
8	0.125809266198656\\
9	0.0661441157130687\\
10	0.0673257264538966\\
11	0.0885092421455029\\
12	0.148238387141577\\
13	0.0264424822400462\\
14	0.0493307701488435\\
15	0.0301654955207687\\
16	0.121010022694884\\
17	0.0593894327667528\\
18	0.0597602435752879\\
19	0.0246716097500183\\
20	0.0385881766524143\\
21	0.10559105712871\\
22	0.125176580847014\\
23	0.111824633010669\\
24	0.0947032635089333\\
25	0.13131443512261\\
26	0.0829528082518841\\
27	0.182097789801394\\
28	0.0578183869993756\\
29	0.116933449465601\\
30	0.144749461428704\\
31	0.126943161133796\\
32	0.235637919801675\\
33	0.101893869058206\\
34	0.0324044300803358\\
35	0.0236404068128193\\
36	0.048194753766114\\
37	0.0648670013227506\\
38	0.0254005396933729\\
39	0.0579483957883275\\
40	0.0504030792397435\\
};
\addlegendentry{RBP-PMBM}
\addplot [color=mycolor2, line width=2.0pt, dashed]
  table[row sep=crcr]{%
1	0.253872309690221\\
2	0.286497782391864\\
3	0.260549355682194\\
4	0.31290178537244\\
5	0.241411775516621\\
6	0.313526155225513\\
7	0.285320100650327\\
8	0.266907101227159\\
9	0.319957166736913\\
10	0.260089706393788\\
11	0.266839977504255\\
12	0.257382662007393\\
13	0.264675740815087\\
14	0.248587376882623\\
15	0.25251865086705\\
16	0.248429780944707\\
17	0.292761299646314\\
18	0.264155482149693\\
19	0.243003655326713\\
20	0.272297839618443\\
21	0.247111310883289\\
22	0.225993574771313\\
23	0.247302216595056\\
24	0.186159102498665\\
25	0.244304959270289\\
26	0.177445404898966\\
27	0.230979511641287\\
28	0.181388395138674\\
29	0.266528825482145\\
30	0.18004473555537\\
31	0.173193737817831\\
32	0.175620053629193\\
33	0.186055582628016\\
34	0.211009745836464\\
35	0.174413915015655\\
36	0.188013466107511\\
37	0.15083920592114\\
38	0.169297676272941\\
39	0.178528251701103\\
40	0.173107178268878\\
};
\addlegendentry{EK-PHD}
\end{axis}

\end{tikzpicture}%
\caption{Comparison of vehicle position estimation among 5 algorithms.}
\label{Fig.positioning_line}
\vspace{-4mm} \end{figure}
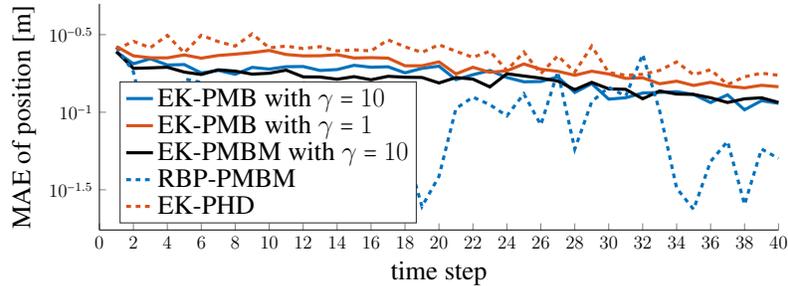

Another main advantage of the proposed SLAM filter is the low computational cost. In order to show this benefit, we measured the average execution time per time step of five SLAM filters, as shown in Table~\ref{tab:cpu_time}. We measured that the proposed EF-PMB SLAM filter with $\gamma=10$ takes 14.4 ms per time step, with the prediction and the update steps costs 0.34 ms and 14.1 ms, respectively, while the RBP-PMBM SLAM filter takes 71865.9 ms per time step, with the prediction and the update steps costs 581.8 ms and 71284.1 ms, respectively. It is obvious that the proposed EK-PMB algorithm is able to reduce the complexity, which is 5000 times faster than the RBP-PMBM SLAM filter approximately. One of the reasons is that the later uses 2000 particles to propagate the non-linearity of the \ac{ue} state and there is an PMBM density presenting the map conditioned on each particle, while the former utilizes the first-order Taylor  extension to approximate the nonlinear model and there is only one PMB(M) over the time. Another reason is that the inverse-CKF is used to create new births and CKF is used to update detected landmarks in the RBP-PMBM algorithm. In the other algorithms, the mean of the newly detected landmarks are directly estimated from the measurements, covariance of the birth components is computed from the Jacobians, and the \ac{ue} and landmark states are jointly updated using the EK filter. The first algorithm takes slightly longer time that the EK-PMB SLAM filter with $\gamma=1$, which is due to more data associations are considered in the update step. The proposed EK-PMBM SLAM filter takes much longer time than the proposed EK-PMB filter, because the PMBM format is kept and the number of global hypotheses rapidly increases. The EK-PHD SLAM filter takes a short time, that is because the low complexity of the PHD filter compared to the PMBM filter, and the hard decisions are taken both on the data association and the type of the landmark. Although the EK-PMB does not has lowest complexity or highest accuracy, online and real-time operation of the filter could still be guaranteed with good accuracy performance. Therefore, it offers a superior overall performance.
\begin{table}[]
    \centering
    \caption{Average computation time in milliseconds of the prediction and update steps of the SLAM filters.}
        \begin{tabular}{ |c|c|c|c| } 
         \hline
         Filter & Prediction & Update & Total \\ \hline 
         EK-PMB with $\gamma=10$ & $0.34$ & $14.1$ & $14.4$
         \\ 
         EK-PMB with $\gamma=1$  & $0.34$ & $11.9$ & $12.3$
         \\ 
         EK-PMBM  & $0.34$ & $41.5$ & $41.9$
         \\ 
         RBP-PMBM  & $581.8$ & $71284.1$ & $71865.9$
         \\ 
         EK-PHD  & $1.2$ & $2.6$ & $3.8$ \\ 
         \hline
        \end{tabular}
        
    \label{tab:cpu_time}
\end{table}

\section{Conclusions} \label{conclusion}
mmWave bistatic sensing is of great relevance to 5G and Beyond 5G systems. When the sensing device is mobile, a SLAM problem needs to be solved to simultaneously localize the sensor and determine the locations of landmarks in the environment. %The particle-based PMBM SLAM filter is a powerful means to solve the SLAM problem, though it comes at a high computational cost, due to both the use of a large number of particles and the rapidly increasing number of data association hypotheses in the PMBM filter. 
In this paper, we have proposed two novel, low-complexity SLAM filters, based on the PMBM and PMB filters, which utilize an EK filter to perform a joint update of the sensor state and the landmark states. An extension to multiple models, relevant for the mmWave bistatic sensing problem, is also introduced. Via simulation results using reasonable mmWave signal parameters, we demonstrate that the proposed filter %is extended  
%To address the challenge of the high computational cost of the RBPF-PMBM SLAM filter, as well as not sacrifice too much 
can attain good mapping and positioning performance, with very low complexity. % we have proposed a novel algorithm for 5G SLAM which is based on the EKF-PMBM filter. 
Our results also demonstrated that the proposed SLAM filter can not only handle mapping and vehicle state estimation simultaneously, but also distinguish the type of landmarks accurately, which is comparable to the performance of the RBP-PMBM SLAM filter. %Meanwhile, the proposed SLAM filter greatly reduces the computational cost with respect to the RBPF-PMBM SLAM filter, saving 0.47 \textperthousand cost. 
The high mapping and positioning performance and the low computational overhead of the proposed EK-PMB(M) SLAM filter are attractive for real-time execution of 5G and Beyond 5G mmWave SLAM algorithms.

Future work will include comparison with additional SLAM filters, the use of ray tracing data to validate the performance under more realistic operating conditions, the inclusion of high-dimensional channel estimation and optimized signal design (precoding and combining) to boost the localization accuracy, batch processing of measurements over a time window, as well as the extension to a multi-UE and multi-BS setup. 

\appendices
\section{Computation of Data Association Metric}
\subsection{Previously Detected Landmark $i$, Hypothesis $j$ Detected Again} \label{app:caseA}
The value of $\rho_{k+1|k+1}^{j,i,p}$ is given by 
\begin{align}
\rho_{k+1|k+1}^{j,i,p}& =\int p_{\text{D}}(\tilde{\boldsymbol{s}}_{k+1}^{j,i})f(\boldsymbol{z}^{p}_{k+1}|\tilde{\boldsymbol{s}}_{k+1}^{j,i})f(\tilde{\boldsymbol{s}}_{k+1}^{j,i})\text{d}\tilde{\boldsymbol{s}}_{k+1}^{j,i}=p_{\text{D},k+1}^{j,i}{\cal N}(\boldsymbol{z}^{p} ; \boldsymbol{h}(\tilde{\boldsymbol{m}}^{j,i}_{k+1|k}),\boldsymbol{S}^{j,i}_{k+1|k})\label{detected4},%\\
\end{align}
where $p_{\text{D}}(\tilde{\boldsymbol{s}}_{k+1}^{j,i})$ is assumed to be a constant over $\tilde{\boldsymbol{s}}_{k+1}^{j,i}$, with the value $p_{\text{D},k+1}^{j,i}= p_{\text{D}}(\tilde{\boldsymbol{s}}_{k+1}^{j,i}=\tilde{\boldsymbol{m}}^{j,i}_{k+1|k})$, $\tilde{\boldsymbol{m}}^{j,i}_{k+1|k}=[\boldsymbol{m}_{k+1|k}^ {\text{T}},(\boldsymbol{u}^{j,i}_{k+1|k})^ {\text{T}}]^{\text{T}}$ is the mean of $\tilde{\boldsymbol{s}}_{k+1}^{j,i}$, which is the joint state of the UE state and $j,i$th landmark state. The innovation covariance  $\boldsymbol{S}_{k+1|k}^{j,i,p}$ is  given by
\begin{align}
    \boldsymbol{S}^{j,i,p}_{k+1|k} = \boldsymbol{H}^{j,i}_{k+1|k} \tilde{\boldsymbol{P}}^{j,i}_{k+1|k} (\boldsymbol{H}^{j,i}_{k+1|k})^{\text{T}}+ \boldsymbol{R}^{p}_{k+1} ,\label{innovation_covariance}
\end{align}
in which $\boldsymbol{H}^{j,i}_{k+1|k}$  represents the Jacobian of $\boldsymbol{h}(\cdot)$ with respect to $\tilde{\boldsymbol{s}}^{j,i}_{k+1}$, evaluated at $\tilde{\boldsymbol{s}}_{k+1}^{j,i}=\tilde{\boldsymbol{m}}^{j,i}_{k+1|k}$,  with elements
\begin{align}
& \boldsymbol{H}^{j,i}_{k|k} =\left.\frac{\partial \boldsymbol{h}(\tilde{\boldsymbol{s}}_{k+1}^{j,i})}{\partial \tilde{\boldsymbol{s}}_{k+1}^{j,i}}\right|_{\tilde{\boldsymbol{s}}^{j,i}_{k+1}=\tilde{\boldsymbol{m}}^{j,i}_{k+1|k}}.
\end{align}
The covariance of $\tilde{\boldsymbol{s}}^{j,i}_{k+1}$ is given by $\tilde{\boldsymbol{P}}^{j,i}_{k+1|k}=\text{blkdiag}(\boldsymbol{P}_{k+1|k},\boldsymbol{C}^{j,i}_{k+1|k})$.

\subsection{Newly Detected Landmark} \label{app:caseC}
The value of $\rho_{\text{B},k+1|k+1}^{p}$ is given by 
\begin{align}
    &\rho_{\text{B},k+1|k+1}^{p}=\int p_{\text{D},k+1}^{p}\eta_{k+1|k} f(\boldsymbol{z}^{p}_{k+1}|\tilde{\boldsymbol{s}}_{k+1}^{p})f(\tilde{\boldsymbol{s}}_{k+1}^{p})\text{d}\tilde{\boldsymbol{s}}_{k+1}^{p}\nonumber\\&\qquad=p_{\text{D},k+1}^{p}\eta_{k+1|k} {\cal N}(\boldsymbol{z}^{p} ; \boldsymbol{h}(\tilde{\boldsymbol{m}}^{p}_{\text{B},k+1|k}),\boldsymbol{S}^{p}_{\text{B},k+1|k})\label{rhonewdetected},
\end{align}
where  $p_{\text{D},k+1}^{p}$ is a constant with  $ p_{\text{D}}(\tilde{\boldsymbol{s}}_{k+1}^{p}=\tilde{\boldsymbol{m}}^{p}_{\text{B},k+1|k})$,  $\tilde{\boldsymbol{m}}^{p}_{\text{B},k+1|k}=[\boldsymbol{m}_{k+1|k}^ {\text{T}},(\boldsymbol{u}^{p}_{\text{B},k+1|k})^{\text{T}}]^ {\text{T}}$ is the mean of $\tilde{\boldsymbol{s}}^{p}_{k+1}$, which is the joint state of the UE state and the newly detected landmark state $\boldsymbol{x}^{p}_{k+1}$. The innovation covariance  $\boldsymbol{S}^{p}_{k+1}$ is given by
\begin{align}
    \boldsymbol{S}^{p}_{\text{B},k+1|k} = \boldsymbol{H}^{p}_{\text{B},k+1|k} \tilde{\boldsymbol{P}}^{p}_{\text{B},k+1|k} (\boldsymbol{H}^{p}_{\text{B},k+1|k})^{\text{T}}+ \boldsymbol{R}_{k+1}^{p} ,\label{innovation_covariance_h}
\end{align}
where $\boldsymbol{H}^{p}_{\text{B},k+1|k}$ is the Jacobian of $\boldsymbol{h}(\tilde{\boldsymbol{s}}^{p}_{k+1})$,  given by 
\begin{align}
    \boldsymbol{H}^{p}_{\text{B},k+1|k}=\left.\frac{\partial \boldsymbol{h}(\tilde{\boldsymbol{s}}^{p}_{k+1})}{\partial \tilde{\boldsymbol{s}}^{p}_{k+1}}\right|_{\tilde{\boldsymbol{s}}^{p}_{k+1}=\tilde{\boldsymbol{m}}^{p}_{\text{B},k+1|k}}. 
\end{align}
The covariance of $\tilde{\boldsymbol{s}}^{p}_{k+1}$ is given by $\tilde{\boldsymbol{P}}^{p}_{\text{B},k+1|k}=\text{blkdiag}(\boldsymbol{P}_{k+1|k},\boldsymbol{C}^{p}_{\text{B},k+1|k})$. Generation of the mean and covariance of a new landmark is explained in Appendix \ref{app:birth}.

\subsection{Birth Generation} \label{app:birth} 
The mean of newly detected landmark $\boldsymbol{u}^{p}_{\text{B},k+1|k+1}$ can be estimated by \cite[Appendix B]{kim20205g}. The covariance of the newly detected landmark can be computed as follows. Consider a joint prior distribution of the sensor state and the newly detected landmark of the form $$\mathcal{N}(\tilde{\boldsymbol{s}}_{k+1};[\boldsymbol{m}^{\mathrm{T}}_{k+1|k},(\boldsymbol{u}^{p}_{\text{B},k+1|k+1})^{\mathrm{T}}]^{\mathrm{T}},\text{blkdiag}(\boldsymbol{P}_{k+1|k},\boldsymbol{C}^{\text{prior}})),$$ with $\boldsymbol{C}^{\text{prior}} \to \infty \boldsymbol{I}$. Then, after applying an EK filter update, the marginal posterior covariance of the landmark %,  
 %$\boldsymbol{C}^{p}_{\text{B},k+1|k}$, 
 is given by 
%\begin{align}
%    \boldsymbol{C}^{p}_{\text{B},k+1|k} = ((\grave{\boldsymbol{H}}^{p}_{\text{B},k+1|k})^{\text{T}} (\tilde{\boldsymbol{S}}^{p}_{\text{B},k+1|k})^{-1} \grave{\boldsymbol{H}}^{p}_{\text{B},k+1|k})^{-1} ,\label{covariance_birth}
%\end{align}
\begin{align}
    & \boldsymbol{C}^{p}_{\text{B},k+1|k+1} =\label{covariance_birth} ((\grave{\boldsymbol{H}}^{p}_{\text{B},k+1|k})^{\text{T}} (\acute{\boldsymbol{H}}^{p}_{\text{B},k+1|k} \boldsymbol{P}_{k+1|k} (\acute{\boldsymbol{H}}^{p}_{\text{B},k+1|k})^{\text{T}}+ \boldsymbol{R}^{p}_{k+1})^{-1} \grave{\boldsymbol{H}}^{p}_{\text{B},k+1|k})^{-1},% \notag 
\end{align}
with % $\tilde{\boldsymbol{S}}^{p}_{\text{B},k+1|k} = \acute{\boldsymbol{H}}^{p}_{\text{B},k+1|k} \boldsymbol{P}_{k+1|k} (\acute{\boldsymbol{H}}^{p}_{\text{B},k+1|k})^{\text{T}}+ \boldsymbol{R}$ considers both uncertainty of of the vehicle state estimate and measurements,
%$\grave{\boldsymbol{H}}^{p}_{\text{B},k+1|k}$ and $\acute{\boldsymbol{H}}^{p}_{\text{B},k+1|k}$ are 
the Jacobians of $\boldsymbol{h}(\cdot)$
\begin{align}
    & \grave{\boldsymbol{H}}^{p}_{\text{B},k+1|k}=%\\& 
    %\quad
    \left.\frac{\partial \boldsymbol{h}(\boldsymbol{s}_{k+1}=\boldsymbol{m}_{k+1|k},\boldsymbol{x}^{p}_{k+1})}{\partial \boldsymbol{x}^{p}_{k+1}}\right|_{\boldsymbol{x}^{p}_{k+1}=\boldsymbol{u}^{p}_{\text{B},k+1|k+1}},\nonumber \\
    & \acute{\boldsymbol{H}}^{p}_{\text{B},k+1|k}=%\\& \quad
    \left.\frac{\partial \boldsymbol{h}(\boldsymbol{s}_{k+1},\boldsymbol{x}^{p}_{k+1}=\boldsymbol{u}^{p}_{\text{B},k+1|k})}{\partial \boldsymbol{s}_{k+1}}\right|_{\boldsymbol{s}_{k+1}=\boldsymbol{m}_{k+1|k}}.\nonumber
\end{align}
%\begin{align}
 %   & \grave{\boldsymbol{H}}^{p}_{\text{B},k+1|k}=\\& \quad\left.\frac{\partial \boldsymbol{h}(\boldsymbol{s}_{k+1|k}=\boldsymbol{m}_{k+1|k},\boldsymbol{x}^{p}_{\text{B},k+1|k})}{\partial \boldsymbol{x}^{p}_{\text{B},k+1|k}}\right|_{\boldsymbol{x}^{p}_{\text{B},k+1|k}=\boldsymbol{u}^{p}_{\text{B},k+1|k+1}},\nonumber \\
  %  & \acute{\boldsymbol{H}}^{p}_{\text{B},k+1|k}=\\& \quad\left.\frac{\partial \boldsymbol{h}(\boldsymbol{s}_{k+1|k},\boldsymbol{x}^{p}_{\text{B},k+1|k}=\boldsymbol{u}^{p}_{\text{B},k+1|k})}{\partial \boldsymbol{s}_{k+1|k}}\right|_{\boldsymbol{s}_{k+1|k}=\boldsymbol{m}_{k+1|k}},\nonumber
%\end{align}
%respectively.

\balance 
\bibliography{IEEEabrv,Bibliography}

% \IEEEtriggeratref{7}
% \IEEEtriggercmd{\enlargethispage{-20cm}}


%\bibliographystyle{IEEEtran}

\end{document}